\documentclass[aps,prl,superscriptaddress,twocolumn,longbibliography]{revtex4-1}

\usepackage{amsfonts}
\usepackage{subfigure}
\usepackage{amsmath}
\usepackage{txfonts}
\usepackage{amssymb}
\usepackage{amsbsy} 
\usepackage{epsfig}
\usepackage{graphicx}

\def\be{\begin{equation}} \def\ee{\end{equation}}
\def\bea{\begin{eqnarray}} \def\eea{\end{eqnarray}}

\def\nn{\nonumber}

\def\bq{{\bf q}}

\def\bk{{\bf k}}

\def\be{{\bf e}}

\def\bA{{\bf A}}

\def\bj{{\bf j}}

\def\la{\langle}
\def\ra{\rangle}

\def\rw{\rightarrow}

\begin{document}

\title{Majorana Zero Modes Protected by Hopf Invariant in Topologically Trivial Superconductors}

\author{Zhongbo Yan}
\affiliation{ Institute for
Advanced Study, Tsinghua University, Beijing, 100084,  China}

\author{Ren Bi}
\affiliation{ Institute for
Advanced Study, Tsinghua University, Beijing, 100084,  China}

\author{Zhong Wang}
\altaffiliation{  wangzhongemail@tsinghua.edu.cn} \affiliation{ Institute for
Advanced Study, Tsinghua University, Beijing, 100084,  China}

\affiliation{Collaborative Innovation Center of Quantum Matter, Beijing, 100871, China }


\begin{abstract}

Majorana zero modes are usually attributed to topological superconductors. We study a class of two-dimensional topologically trivial superconductors without chiral edge modes, which nevertheless host robust Majorana zero modes in topological defects. The construction of this minimal single-band model is facilitated by the Hopf map and the Hopf invariant. This work will stimulate investigations of Majorana zero modes in superconductors in the topologically trivial regime.

\end{abstract}


\maketitle

Majorana zero modes (MZMs) or Majorana bound states are exotic excitations predicted to exist in the vortex cores\cite{read2000,volovik1999fermion} of two-dimensional (2D) topological superconductors\cite{hasan2010,qi2011,Bansil2016, bernevig2013topological,shen2013topological} and at the ends of 1D topological superconductors\cite{kitaev2001unpaired}. Spatially separated MZMs give rise to degenerate ground states, which encode qubits immune to local dechoerence\cite{kitaev2001unpaired,nayak2008}. Furthermore, unitary transformations among the ground states can be implemented by braiding\cite{moore1991nonabelions,wen1991a,ivanov2001,nayak1996,sarma2005} or measurements\cite{vijay2016,bonderson2008} of these modes, indicating that such qubits may become building blocks in topological quantum computation and information\cite{aasen2016,karzig2016, karzig2016scalable,heck2012coulomb,landau2016,deng2013fault}. Therefore,
MZMs have been vigorously pursued in condensed matter physics\cite{alicea2012new,Beenakker2013,stanescu2013majorana, leijnse2012introduction,Elliott2015,sarma2015majorana,sato2016majorana}.

There have been a great variety of proposals for topological superconductors, including 2D semiconductor heterostructures\cite{sau2010,alicea2010}, topological insulator-superconductor proximity\cite{fu2007c,qi2010chiral,Chung2011, law2009majorana,akhmerov2009}, 1D spin-orbit-coupled quantum wires\cite{oreg2010helical,lutchyn2010,alicea2011non,lutchyn2011, stanescu2011majorana,potter2010multchannel,Rainis2013,Prada2012,sarma2012splitting}, spiral magnetic chains on superconductors\cite{choy2011,martin2012,nadj2013proposal,Klinovaja2013,vazifeh2013}, Shockley mechanism\cite{wimmer2010,Deng2015proposal}, and cold atom systems in 2D\cite{sato2009non,zhang2008px,tewari2007quantum,liu2014realization} and 1D\cite{Jiang2011,diehl2011topology}, etc. Experimentally, suggestive signatures of MZMs in both 1D\cite{mourik2012signatures, nadj2014observation,rokhinson2012fractional,deng2012anomalous, das2012zero,finck2013,churchill2013superconductor, albrecht2016exponential,Deng2016Majorana,franz2013majorana,pawlak2015probing} and 2D\cite{Xu2015experimental, Sun2016Majorana,lv2016,Wang2012coexistence,wang2016topological,he2016chiral}  topological superconductors have been found.

It is often implicitly assumed that topological superconductivity is a prerequisite for MZMs, accordingly, the chiral edge states go hand in hand with the vortex zero modes in 2D superconductors. In this Letter we show that certain topological defects\cite{teo2010,Chiu2015RMP,shiozaki2014,lee2007,qi2008b,ran2009one} in 2D topologically \emph{trivial} superconductors can support robust MZMs. Somewhat surprisingly, single-band superconductors suffice this purpose. The model Hamiltonian is related to the \emph{Hopf maps}, which originally refer to nontrivial mappings from a 3D sphere $S^3$ to a 2D sphere $S^2$, characterized by the integer Hopf invariant\cite{wilczek1983,nakahara2003}. Mappings from a 3D torus $T^3$ to $S^2$ inherit the nontrivial topology from the mappings $S^3\rw S^2$. The Hopf invariant has found interesting applications in nonlinear $\sigma$ models and spin systems\cite{wilczek1983,fradkin2013}, Hopf insulators\cite{moore2008topological,deng2013hopf,deng2015systematic, deng2016probe,kennedy2016,liu2016symmetry}, liquid crystals\cite{Ackerman2017}, and quench dynamics of Chern insulators\cite{wang2016measuring,flaschner2016observation}.

\begin{figure}
\subfigure{\includegraphics[width=4.3cm, height=4cm]{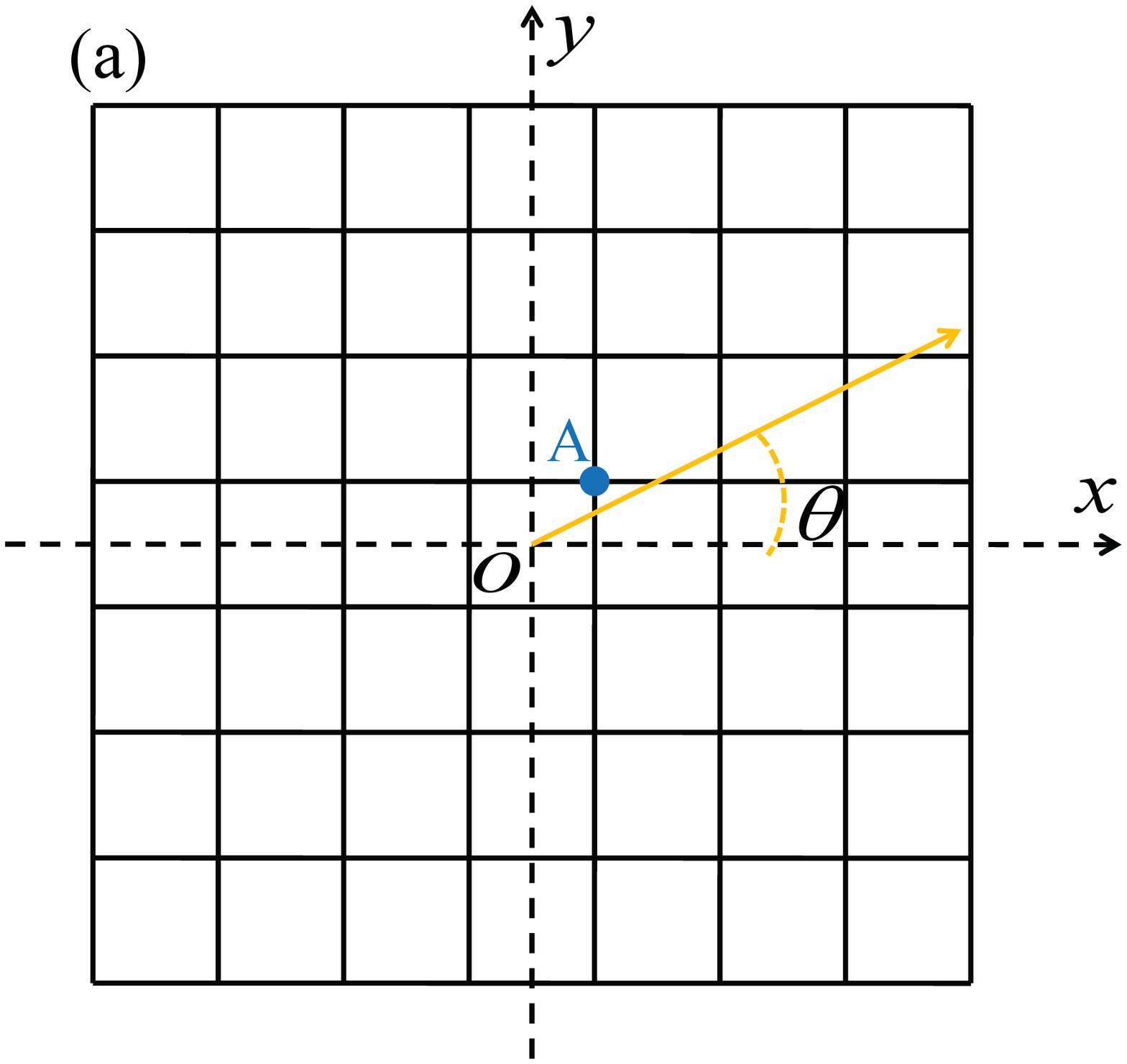}}
\subfigure{\includegraphics[width=4.2cm, height=4.0cm]{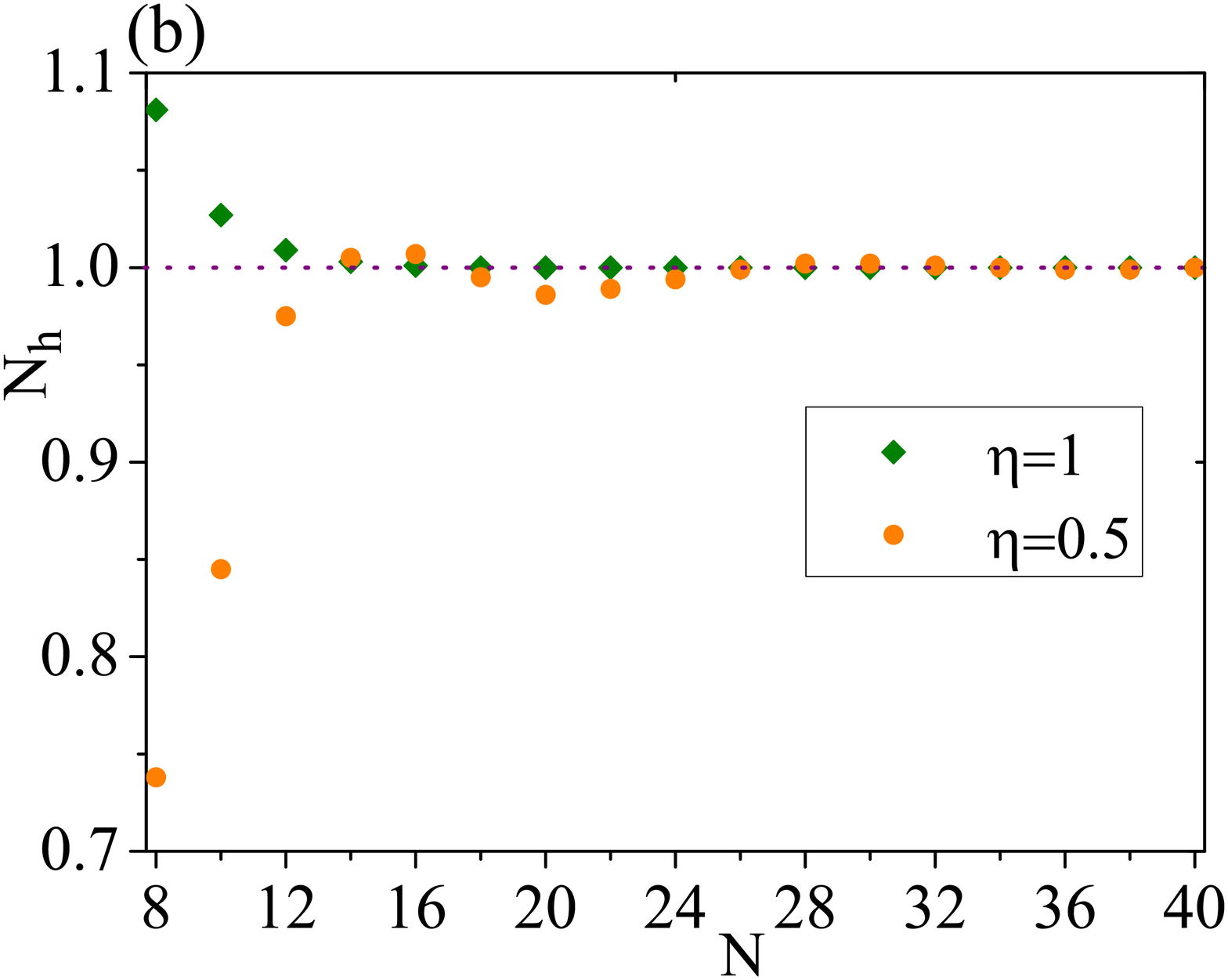}}
\caption{ (a) Sketch. The Hamiltonian varies as a function of $\theta$, creating a defect at $O=(x_0,y_0)$. In the polar coordinate, $\theta\equiv \arctan(y-y_0)/(x-x_0)$. (b) The Hopf invariant for $\eta=1$ and $\eta=0.5$ ($n=1$) in discretized-zone calculation, with $N^3$ grid points in the $(\bk,\theta)$ space. As the grid becomes finer, $N_h$ converges rapidly to $1$. }  \label{sketch}
\end{figure}

Our model describes topologically trivial superconductors with zero Chern number and no chiral edge state. Nevertheless, a topological point defect is characterized by a Hopf invariant defined in the $(k_x,k_y,\theta)$ space, where $k_x,k_y$ are crystal momenta and $\theta$ is the polar angle\footnote{Converting a momentum variable to an angle variable was
adopted in Refs.\cite{qi2008}}(Fig.\ref{sketch}a). The parity (even/odd) of Hopf invariant determines the presence (absence) of robust MZMs, though the superconductor for every fixed $\theta$ is topologically trivial.
Stimulated by this mechanism, which significantly differs from the magnetic-vortex origin of zero mode in topological $p$-wave superconductor\cite{read2000,volovik1999fermion}, we design trivial-superconductor-based (and vortex-free) T-junctions harboring MZMs.

{\it Zero modes.--}Before studying topological defects, we consider spatially uniform 2D single-band Bogoliubov-de Gennes (BdG) Hamiltonians parameterized by $\lambda$:
\bea H(\bk,\lambda)=\left(
  \begin{array}{cc}
    \xi_\bk(\lambda) &  \Delta_\bk(\lambda) \\
    \Delta_\bk^*(\lambda) & -\xi_{-\bk}(\lambda)  \\
  \end{array}
\right) \eea where $\bk=(k_x,k_y)$, $\xi_\bk=E_\bk-\mu$, $E_\bk$ and $\mu$ is the energy and chemical potential, respectively, and $\Delta_\bk$ is the Cooper pairing. It describes single-band spinless (or spin-fully-polarized) superconductors. This Hamiltonian can be written in terms of the Pauli matrices $\tau_i$ as
\bea H(\bk,\lambda)=\sum_{i=x,y,z}d_i(\bk,\lambda)\tau_i,\label{tau} \eea with $d_x={\rm Re}\Delta_\bk$, $d_y=-{\rm Im}\Delta_\bk$, $d_z=\xi_{\bk}$ (we have $\xi_\bk=\xi_{-\bk}$ in our model).
For reason to become clear shortly, we take \bea d_{i}=z^{\dag}\tau_{i}z,\eea where
$z=(z_1,z_2)^{T}$ and
\begin{eqnarray}
z_1&=&\sin k_{x}+i\sin k_{y},\nonumber\\
z_2&=&\sin\lambda+i(\cos k_{x}+\cos k_{y}+\cos\lambda-m_0),\label{z}
\end{eqnarray}  with $m_0=\frac{3}{2}$. We can check that $\Delta_{-\bk}=-\Delta_\bk$ and $\Delta_{(-k_y,k_x)}=i\Delta_{(k_x,k_y)}$, thus the pairing is $p$-wave.  Given Eq.(\ref{z}), the pairing $\Delta_\bk$ is of the same order as the hopping $\xi_\bk$. To describe weakly-pairing superconductors, one may consider \bea H_\eta(\bk,\lambda)=\eta(d_x\tau_x+d_y\tau_y)+d_z\tau_z, \label{eta} \eea with a small but nonzero $\eta$. Nevertheless, tuning the value of $\eta$ does not close the energy gap, hence it does not qualitatively change the results. Thus we will simply take $\eta=1$ below. The mathematical form of Eq.(\ref{z}) has been introduced for 3D Hopf insulators\cite{moore2008topological,deng2013hopf,deng2015systematic, deng2016probe,kennedy2016,liu2016symmetry}, with $\lambda$ replaced by the third momentum $k_z$. The physical system we will study is nevertheless not directly related to Hopf insulator.

The familiar Chern number\cite{thouless1982,read2000,qi2010chiral} $C$ that characterizes 2D topological superconductors can be obtained by a straightforward numerical calculation, which yields $C(\lambda)=0$ for every $\lambda$.

A topological defect can be generated if the parameter $\lambda$ depends on spatial coordinates, in a manner that the configuration cannot be smoothly deformed to a spatially uniform one. Let us focus on the defects with $\lambda$ depending on the polar angle $\theta$ (Fig.\ref{sketch}a) as \bea \lambda=n\theta, \label{defect} \eea where $n$ is an integer. These configurations are topologically nontrivial due to a nonzero Hopf invariant, as we now explain. The unit vector ${\bf \hat{d}}(\bk,\theta)\equiv \frac{1}{\sqrt{\eta^2(d_x^2+d_y^2)+d_z^2}} (\eta d_x,\eta d_y,d_z)$ maps the 3D torus $T^3$ ($k_x,k_y,\theta$ are defined modulo $2\pi$) to the 2D unit sphere $S^2$. For nonzero $n$, the inverse-image circles of two points on $S^2$ are linked\cite{supplemental}. To quantify such linking, the Hopf invariant can be defined\cite{wilczek1983,moore2008topological}: $N_h= -\frac{1}{4\pi^2}\int d\theta d^2k\epsilon^{\mu\nu\rho}a_\mu \partial_\nu a_\rho$, where the integrating range is the Brillouin zone for $\bk$ and $[0,2\pi]$ for $\theta$, $a_\mu=-i\la\psi(\bk,\theta)|\partial_\mu|\psi(\bk,\theta)\ra$, with $\mu,\nu,\rho=k_x,k_y,\theta$, and $|\psi\ra$ is the negative-energy eigenfunction of $H_\eta(\bk,\theta)$. Alternatively, we can define $A_\mu=a_\mu/2\pi$, $j^\mu=\epsilon^{\mu\nu\rho}\partial_\nu A_\rho=(1/8\pi)\epsilon^{\mu\nu\rho}\hat{{\bf d}}\cdot(\partial_\nu \hat{{\bf d}}\times\partial_\rho \hat{{\bf d}})$, then\cite{wilczek1983,moore2008topological} \bea N_h=-\int d^2kd\theta\,\bj\cdot\bA. \label{hopf-def} \eea It can be calculated numerically by discretizing the Brillouin zone\cite{supplemental}.  The numerical result for $n=1$ is shown in Fig.\ref{sketch}b. More generally, we have $N_h=n$. We will call the topological defects defined by Eq.(\ref{defect}) as \emph{Hopf defects}.

\begin{figure}
\subfigure{\includegraphics[width=8cm, height=5.4cm]{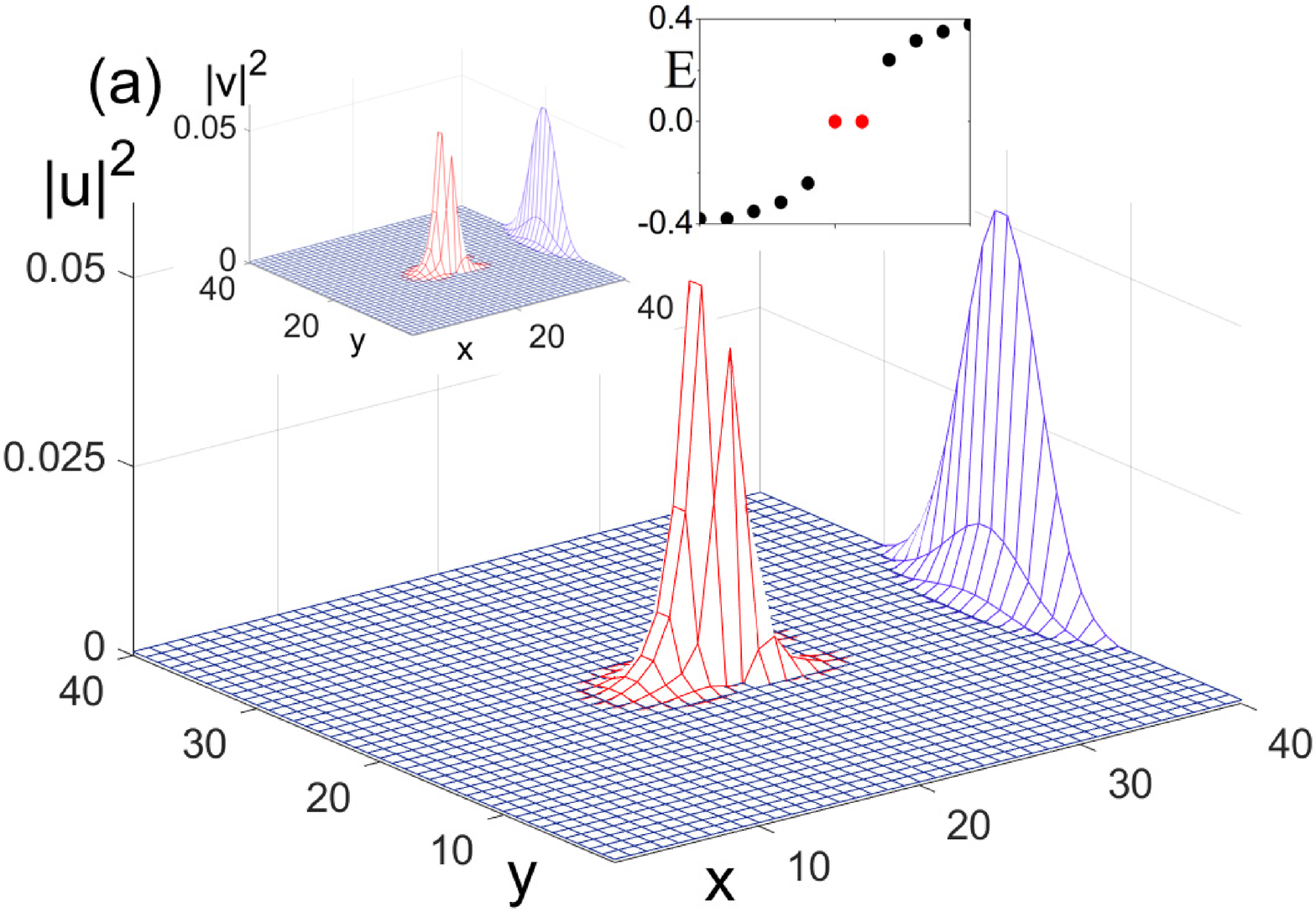}}
\subfigure{\includegraphics[width=8cm, height=5.4cm]{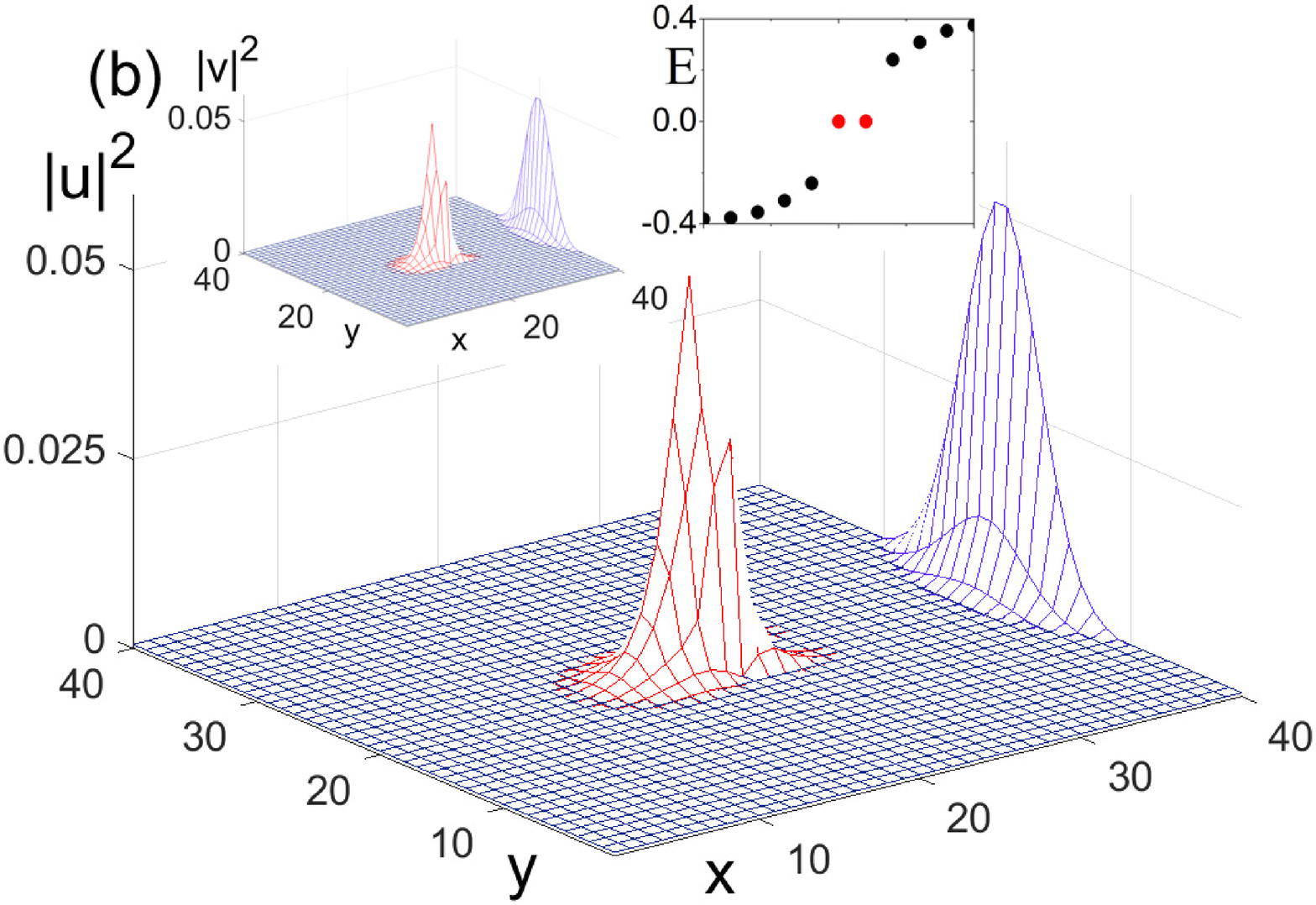}}
\caption{ (a) Profiles of two MZMs in a square sample with a Hopf defect ($n=1$). The main figure shows the particle component $u$, and the left inset shows the hole component $v$. The right inset shows several energies close to 0, with two zero energies colored in red. (b) is the same as (a) except that an impurity potential at the single site $A$ (indicated in Fig.\ref{sketch}a) is added. }  \label{mode}
\end{figure}

To obtain energy spectra, we Fourier-transform the BdG Hamiltonian to real-space lattice, then numerically solve the Hamiltonian\cite{supplemental}. For $n=1$, two zero modes are found  (Fig.\ref{mode}a), one of which is sharply localized around the defect, the other is localized at the sample boundary. The profile  of particle component ($\tau_z=1$ component, denoted by $u$) and hole component ($\tau_z=-1$ component, denoted by $v$) is shown in the main figure and the inset, respectively. It is apparent that the zero modes are equal-weight superpositions of particle and hole components, which is a feature of MZMs (inspection of the wavefunction confirms that $u=v^*$).

To check the robustness of MZMs, we add an impurity potential $Uc^\dag_A c_A$ ($c$ is the fermion operator) at a single site $A$ (specified in Fig.\ref{sketch}a), which amounts to adding a $U\tau_z$ term at site $A$ in the real-space BdG Hamiltonian. The numerical result for $U=1.0$ is shown in Fig.\ref{mode}b. The energies of MZMs remain pinned to $E=0$, though the mode profile is changed compared to Fig.\ref{mode}a.

For $n=2$, we find two localized modes in the defect and two at the boundary (near $\theta=0$ and $\theta=\pi$). Unlike the $n=1$ case, the energies of defect modes are not pinned to zero. Higher $n$'s are also calculated, and the results support the conclusion that there is a single robust MZM for odd-integer $n$, and no robust MZM for even-integer $n$, therefore, it is the Z$_2$ Hopf invariant (even/odd) that determines the existence of MZM in the defect. One may notice that the Hopf invariant takes the form of a Chern-Simons invariant\cite{teo2010,qi2008}, which is not accidental, because the latter is indeed a general topological invariant,
nevertheless, it has been applied\cite{teo2010,Chiu2015RMP} only to topologically \emph{nontrivial} superconductors, for which it is
just the product of the Chern number and the vorticity of pairing phase. Our model shows that nonzero Chern number is not a necessary condition for MZM.

\begin{figure}
\subfigure{\includegraphics[width=4.3cm, height=4cm]{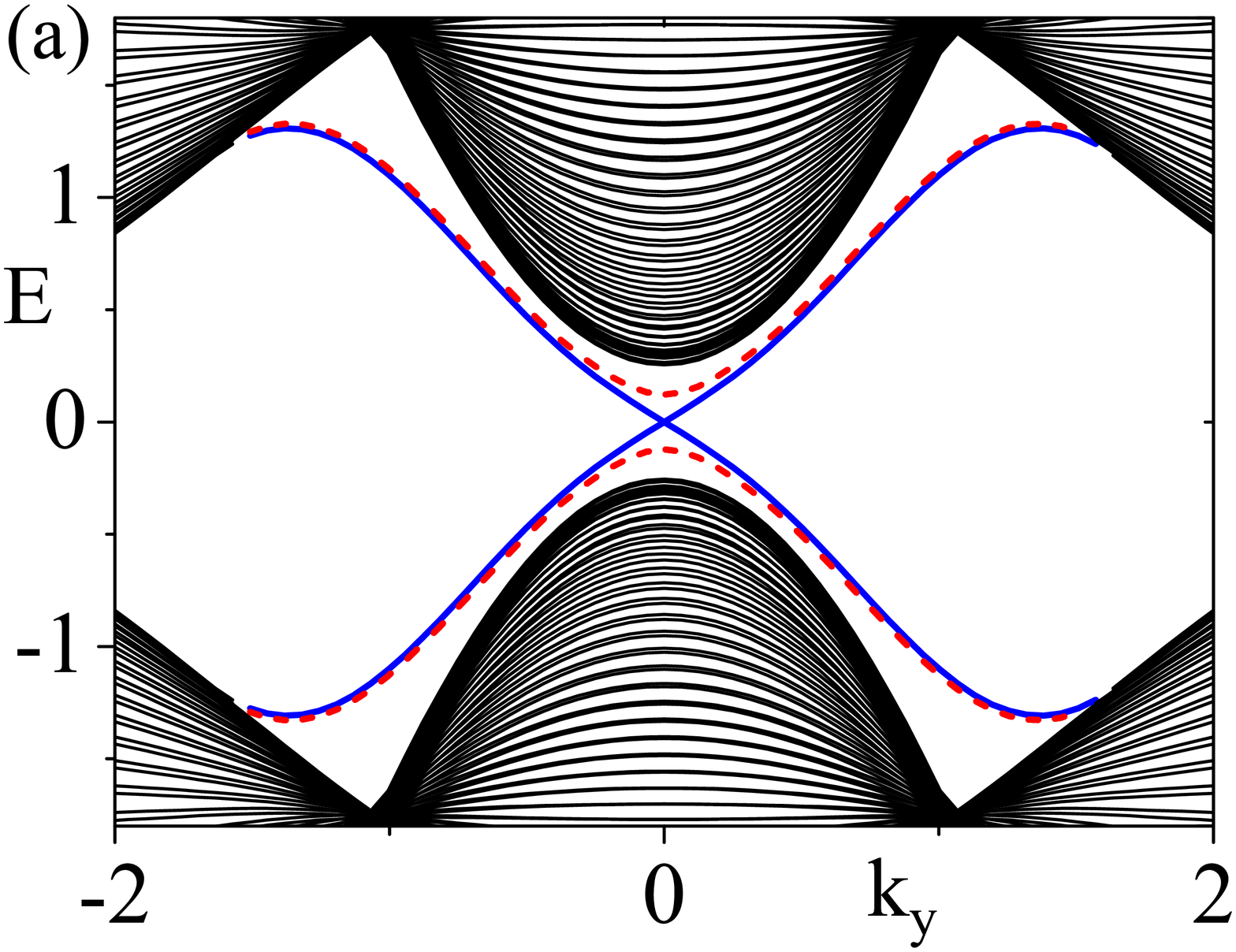}}
\subfigure{\includegraphics[width=4.25cm, height=3.8cm]{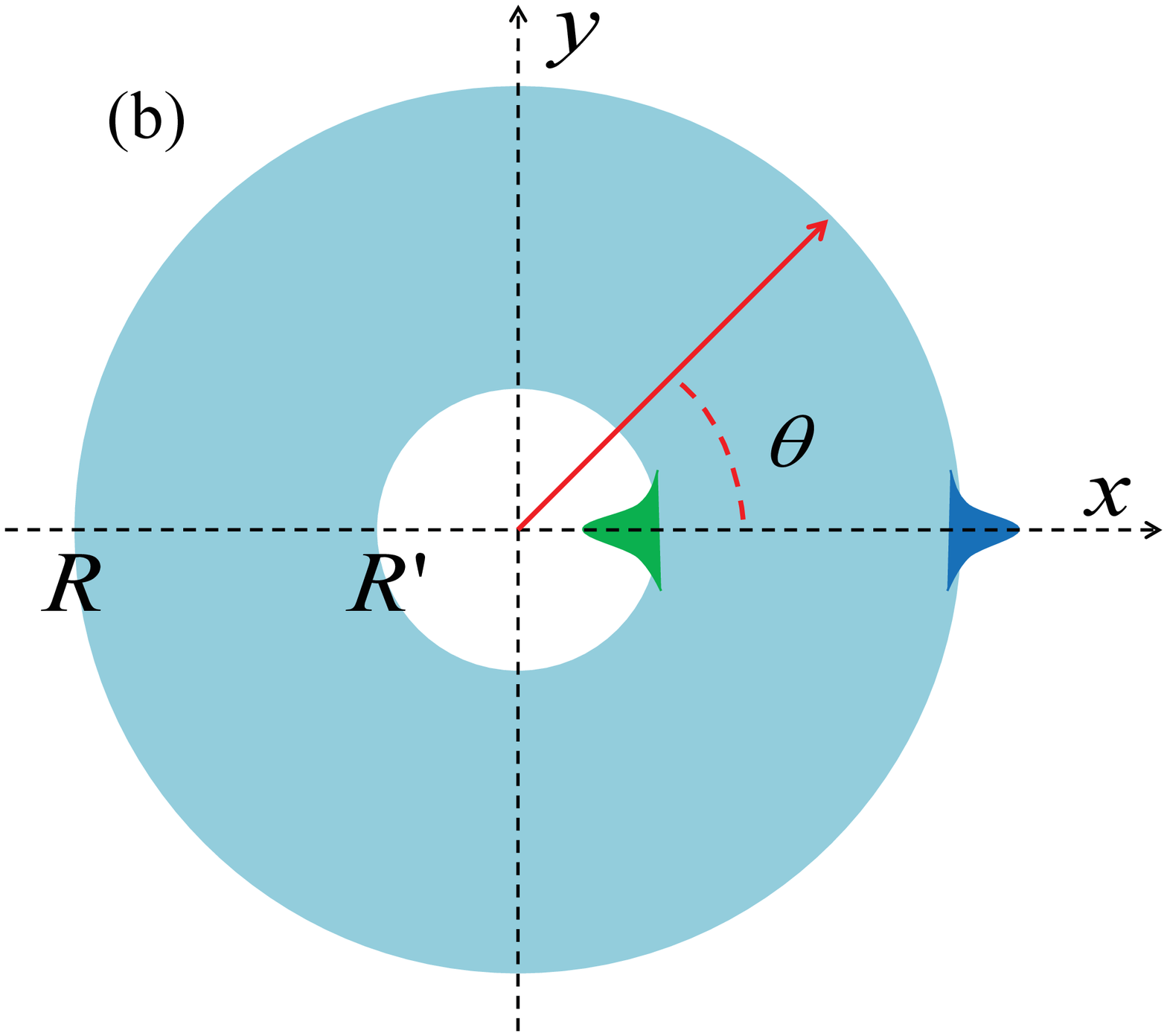}}
\caption{ (a) The energy bands for a ribbon with size $L_x\times L_y=40\times\infty$; $\lambda=0$ (solid curves). The two dashed curves show the gapped edge modes for $\lambda=\pi/20$ as a comparison. Each edge-mode band is doubly degenerate because a ribbon has two boundaries (ignoring a small splitting that exponentially decays as a function of $L_x$). (b) A large hollow disk with two MZMs illustrated. The inner MZM persists as the inner radius $R'\rw 0$, evolving to the defect MZM protected by Hopf invariant. }  \label{disk}
\end{figure}

{\it Edge theory.--}It is desirable to have an intuitive understanding of the MZM from the perspective of edge theory, which, as we will show, differs significantly from that of the chiral topological superconductor\cite{read2000,fendley2007,stone2004}. First, we numerically solved the edge states of open-boundary systems for various values of $\lambda$, and found that gapless edge modes exist only for $\lambda=0$. In Fig.\ref{disk}a, we show the energy bands for a ribbon along $y$ direction. The gapless edge modes for $\lambda=0$ are shown as the solid blue lines. They are non-chiral, and are immediately gapped out when $\lambda$ is tuned away from 0 (edge modes of $\lambda=\pi/20$ are shown in dashed curves), in other words, the edge modes are not topologically robust. This is consistent with the vanishing of Chern number.

This numerical observation is confirmed by analytic solutions. We consider a semi-infinite geometry with the sample occupying $x<0$ region, $k_y$ being a good quantum number. For $\lambda=0$, we obtain two degenerate edge modes at $k_y=0$, both of which are eigenfunctions of $\tau_x$ with eigenvalue $-1$\cite{supplemental}, thus they are equal-weight superpositions of particle and hole components. We introduce Pauli matrices $\sigma_{x,y,z}$ (unrelated to the $\tau_{x,y,z}$ matrices) in this two-dimensional space, so that the two eigenfunctions have $\sigma_z=\pm 1$, respectively. Including small $k_y$ and $\lambda$ as perturbations, we derive an effective theory\cite{supplemental}:
\bea H_{\rm eff}(k_y,\lambda)=-vk_y\sigma_x+ M\lambda\sigma_y,  \label{heff} \eea where the effective parameters $v,M$ are found to be both $3/4$ in our specific model\cite{supplemental}. Thus the edge-state spectra are $E_\pm(k_y)=\pm\sqrt{v^2k_y^2+M^2\lambda^2}$. It is immediately clear that the edge states become gapped when $\lambda$ moves away from 0, which is consistent with the numerical finding in Fig.\ref{disk}a. As a comparison, we note that the edge spectrum of a chiral topological superconductor\cite{read2000},  $E(k_y)=vk_y$, cannot be gapped out.

Based on this effective edge theory, we proceed to study a hollow disk with polar-angle-dependent parameter, $\lambda=\theta$ (Fig.\ref{disk}b). We are only concerned with low-energy modes, therefore, we focus on the neighborhood of $\theta=0$. Suppose that both the outer and inner radiuses $R,R'$ are large. On the outer boundary, we have $\lambda=\theta=y/R$, thus the edge state spectra are given by solving $H_{\rm eff}(k_y\rw -i\partial_y,\lambda\rw y/R)\psi=E\psi$. More explicitly, it reads
\bea [iv\sigma_x\partial_y+(M/R)y\sigma_y]\psi=E\psi, \label{hy} \eea which squares to $(-v^2\partial_y^2+M^2y^2/R^2-vM/R\sigma_z)\psi=E^2\psi$. This equation resembles the Schr\"odinger equation of harmonic oscillators, though $E$ is replaced by $E^2$, and there is a crucial additional $-vM/R\sigma_z$ term. The eigenfunctions are $\sigma_z$-eigenvectors  (eigenvalues are denoted by $s_z=\pm 1$), with energies given by \bea E_{\rm outer}^2(s_z,n)=2(n+1/2)vM/R-s_zvM/R,  \eea where $n=0,1,\cdots$. There is a MZM in the $s_z=+1$ sector, with $n=0$, which is illustrated as the blue bump in Fig.\ref{disk}b. Since this mode is the eigenfunction of $\tau_x$, it is an equal-weight superposition of particle and hole components.

For a semi-infinite geometry with sample occupying the $x>0$ region, the effective edge theory is almost the same as Eq.(\ref{heff}), except that the sign of the first term reversed\cite{supplemental}.
On the inner boundary of the hollow disk (again near $\theta=0$), we have $\lambda=\theta=y/R'$, thus the edge-mode spectrum can be obtained from $[-iv\sigma_x\partial_y +(M/R')y\sigma_y]\psi=E\psi$, analogous to Eq.(\ref{hy}). The energies are given by \bea E^2_{\rm inner}(s_z,n)=2(n+1/2)vM/R'+s_zvM/R',  \eea which features a MZM in the $s_z=-1$ sector. The zero-mode wavefunction is \bea \psi_{\rm inner}\sim\exp(-My^2/2vR')|s_z=-1\ra, \eea which is exponentially localized near $y=0$, namely $\theta=0$ (illustrated by the green bump in Fig.\ref{disk}b). All nonzero energies grow as $1/R'$ as $R'$ is decreased, while the MZM remains at zero energy, evolving to the defect mode shown in Fig.\ref{mode}. For a hollow disk with $\lambda=2\theta$, there are two MZMs on the inner boundary for large $R'$, near $\theta=0$ and $\theta=\pi$, respectively. Shrinking $R'$ causes overlapping between them, which splits the two zero energies to nonzero values. This is consistent with the absence of MZM in the $n=2$ defect.

It is useful to compare our systems with the chiral topological superconductor, for which a magnetic vortex with $\pi$-flux hosts a MZM\cite{read2000,volovik1999fermion,fu2007c,tewari2007index}. In a hollow-disk geometry, this MZM comes from the chiral edge states on the boundary circle\cite{read2000,fendley2007,stone2004}. The MZM  wavefunction is evenly distributed on the circle, which implies its sensitiveness to the magnetic flux. In contrast to this picture, the MZM in our model is not derived from chiral edge state, which is simply absent here, moreover, the MZM is exponentially localized near $\theta=0$ (Fig.\ref{disk}b), thus it is insensitive if a magnetic flux is inserted.

\begin{figure}
\subfigure{\includegraphics[width=8cm, height=4.8cm]{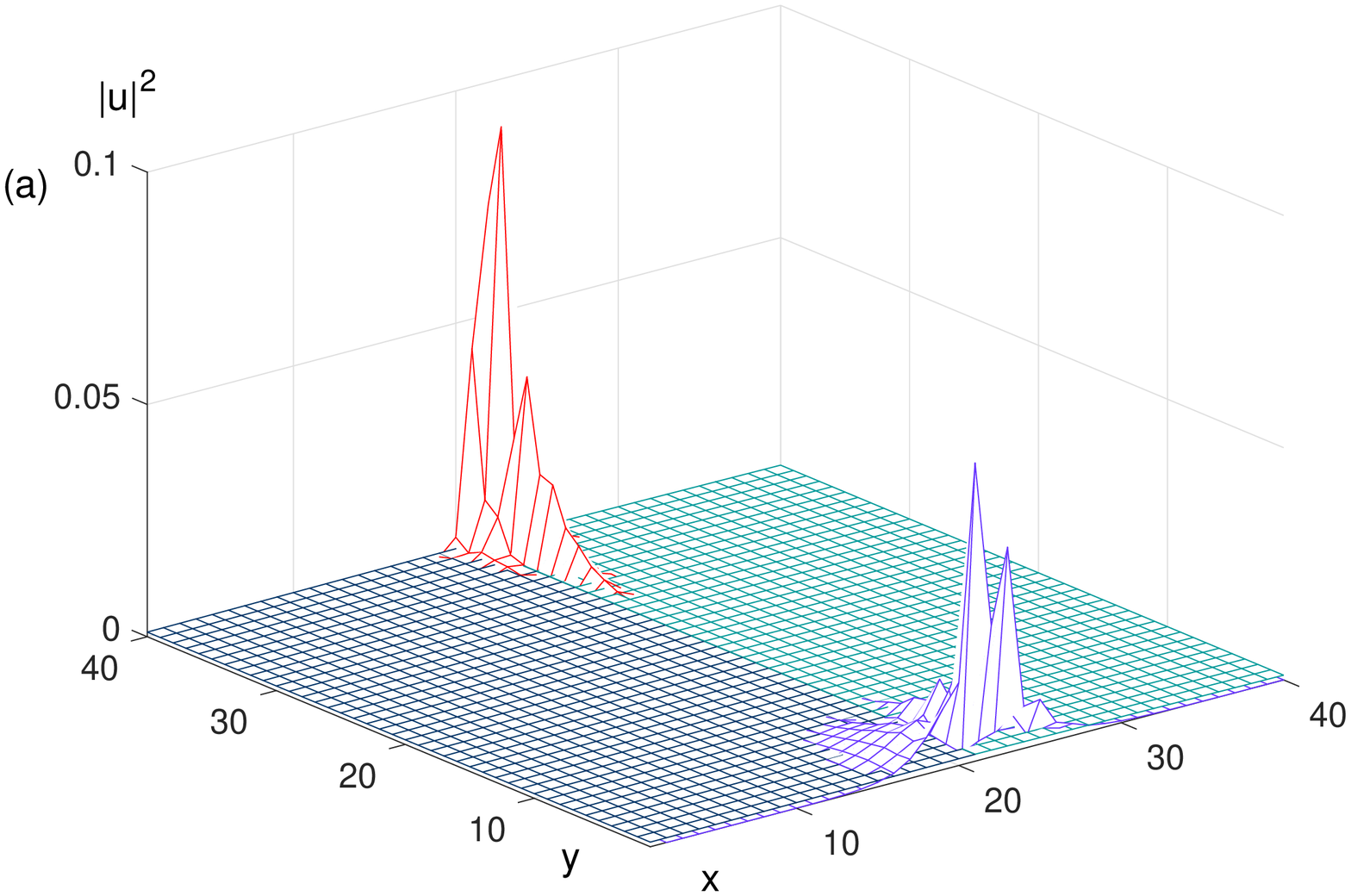}}
\subfigure{\includegraphics[width=8cm, height=4.8cm]{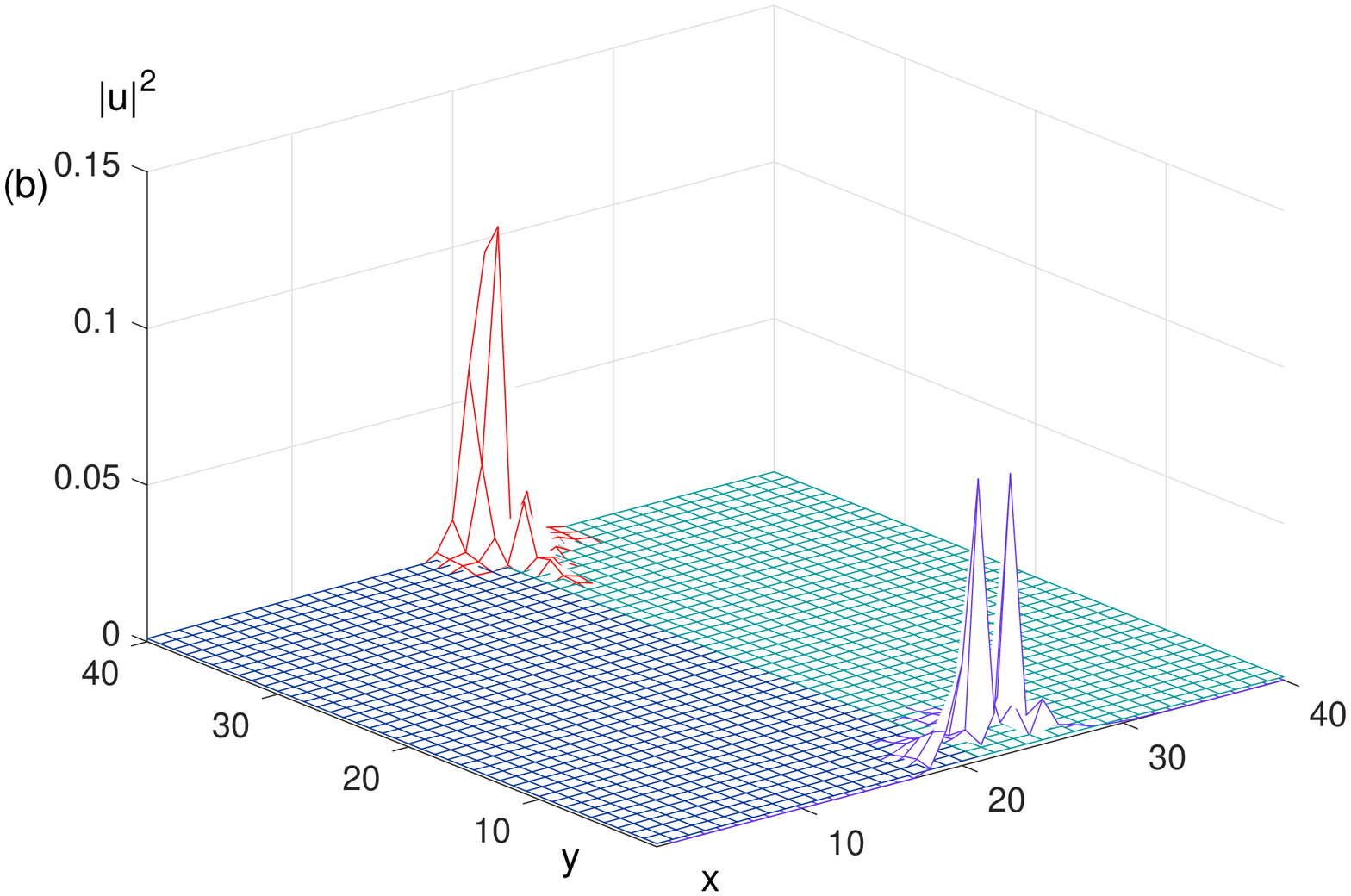}}
\subfigure{\includegraphics[width=4.1cm, height=3.4cm]{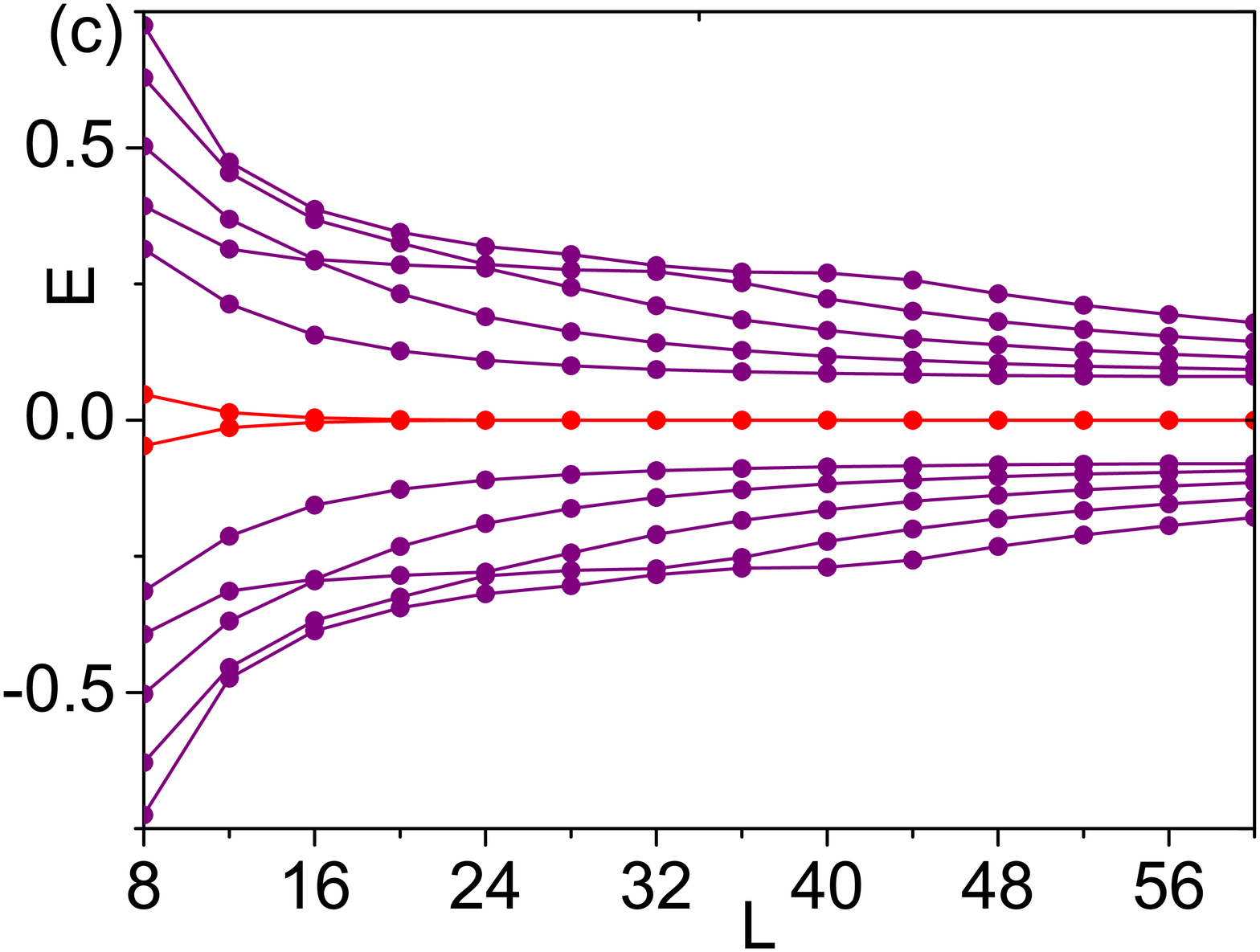}}
\subfigure{\includegraphics[width=4.1cm, height=3.4cm]{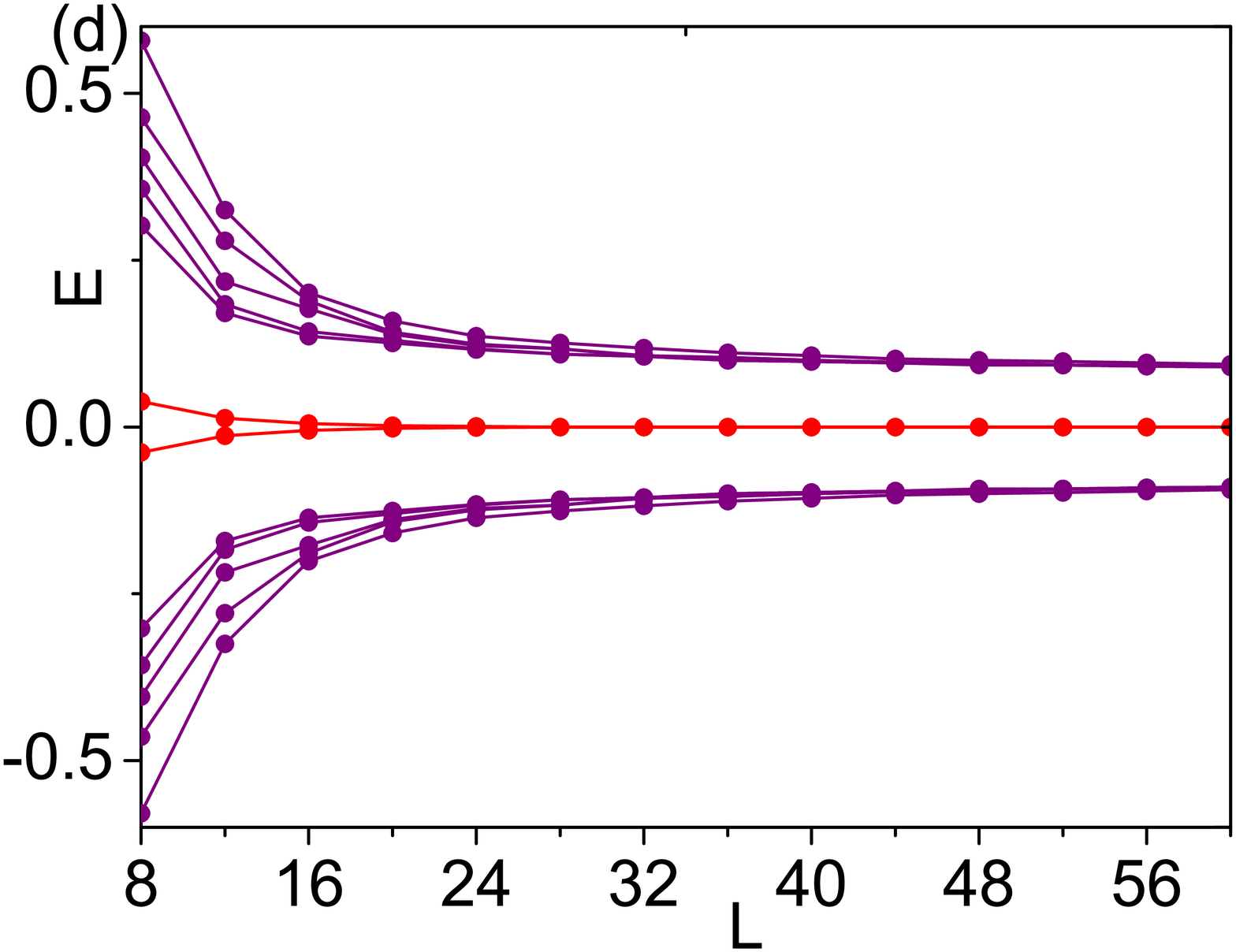}}
\caption{ (a) Superconductor-superconductor-vacuum T-junctions. The system size is $L^2$, with $L=40$. The $x>L/2$ and $x\leq L/2$ region, shown in different colors, is described by Eq.(\ref{tau}) with $\lambda$ taking $\lambda_1=0.1\pi$ and $\lambda_2=\pi$, respectively. The peaks are the profiles of the two MZMs localized around the two T-junctions (each T-junction hosts one mode). (b) Superconductor-insulator-vacuum T-junctions. The parameters are the same as (a) except that $d_x$ and $d_y$ are tuned to 0 in the $\lambda_2$ region, so that $\lambda_2$ region is an insulator. (c) and (d) shows 12 energies closest to 0, for system (a) and (b), respectively.  }  \label{trijunction}
\end{figure}

T-{\it junctions.--}So far, we have only studied configurations with $\lambda$ continuously varied. It is conceivable that the smooth Hopf defect defined by Eq.(\ref{defect}) can be imitated by a discontinuous one, for instance, we may consider a T-junction:
\bea
\lambda(\theta)=\left\{\begin{array}{cc}\lambda_1,&\theta\in [0,\pi/2]\\ \lambda_2,&\theta\in[\pi/2,\pi]\\ \lambda_3,&\theta\in[\pi,2\pi].\end{array}\right.
\eea  $\lambda_{1,2,3}$ being three unequal constants. Such T-junctions will presumably be easier to realize than configurations with $\lambda$ smoothly varying in space.

We will study the simpler superconductor-superconductor-vacuum T-junction by replacing the $\lambda_3$-region by the vacuum. Since the value of $\lambda$ in the vacuum is not well defined, this replacement is not fully justified in advance. Nevertheless, the numerical results thus obtained indicate that MZMs do exist in such T-junctions, as shown in Fig.\ref{trijunction}a for $\lambda_1=0.1\pi$, $\lambda_2=\pi$. There is certain arbitrariness in choosing the hopping at the boundary between the $\lambda_{1,2}$ regions, for which we keep only the nearest-neighbor hopping (discarding the next-nearest-neighbor hopping and the pairing)\footnote{Keeping only the nearest-neighbor hopping means that (in real space) $H_{{\bf r}{\bf s}}=(1.5-\cos\lambda)\tau_z$ for the boundary sites ${\bf r}=(L/2,y)$ and ${\bf s}=(L/2+1,y)$, where we take $\lambda$ as $\lambda_2=\pi$ (Another choice can be $\lambda=\lambda_1$. The choice is not unique).}.  The energy eigenvalues near zero are shown in Fig.\ref{trijunction}c (we show 12 of them), from which it is clear that the zero-mode levels are separate from all other energy levels by a finite gap in the $L\rw\infty$ limit.

We have also studied superconductor-insulator-vacuum T-junctions. To this end, we  consider the $H_\eta$ in Eq.(\ref{eta}), in which taking $\eta=0$ amounts to removing the Cooper pairing.
We notice that $H_{\eta=0}(\bk,\lambda=\pi)$ describes an insulator without any Fermi surface [In contrast, $H_{\eta=0}(\bk,\lambda=0.1\pi)$ describes a metal]. Now we can design superconductor-insulator-vacuum T-junctions by taking $\eta=1$ in the $\lambda_1=0.1\pi$ region, and $\eta=0$ in the $\lambda_2=\pi$ region. We find one MZM for each T-junction (shown in Fig.\ref{trijunction}b), and the energy gap between the zero-mode levels and other energy levels is apparent in Fig.\ref{trijunction}d. We emphasize that each region by itself is topologically trivial, in particular, the superconductor (with $\lambda=0.1\pi$) is a topologically trivial one without gapless edge state.

\emph{Conclusions.--}We have investigated the intriguing possibility of creating MZMs in 2D topologically trivial superconductors. The Hopf defect is constructed as a minimal model for this purpose. Furthermore, we studied the more accessible T-junctions constructed from topologically trivial superconductors. Hopefully, the trivial-superconductor-based approach will broaden the scope of searching MZMs in various superconductors. In particular, absence of chiral Majorana edge state in a 2D superconducting sample does not necessarily imply absence of robust MZM in its point defects.

We conclude with several remarks. First, we have focused on a single-band model, while many materials have multi-bands. We emphasize that the MZMs found here are nevertheless robust to small mixing with other bands, because a single localized MZM cannot move away from zero energy, as required by the intrinsic particle-hole symmetry of the BdG Hamiltonian\cite{hasan2010}. Second, we have taken a simple model BdG Hamiltonians as our starting point (like Ref.\cite{read2000}). More realistic Hamiltonians should be adopted when dealing with real materials, for instance,  the semiconductor-superconductor heterostructures\cite{sau2010,alicea2010}, for which our theory implies that robust MZMs can exist in certain defects (e.g. judiciously constructed T-junctions), without requiring the uniform system being tuned to the topologically nontrivial regime. This will be left for future works.

\emph{Acknowledgements.--}We would like to thank Suk Bum Chung for  helpful suggestions on the manuscript. This work is supported by NSFC (No. 11674189). Z.Y. is supported in part by China Postdoctoral Science Foundation (No. 2016M590082).

\bibliography{dirac}


\newpage

{\bf Supplemental Material}


\section{I. Explicit expressions of $d_{x,y,z}$}

In the main article, we have taken $d_{i}=z^{\dag}\tau_{i}z$  with $z=(z_1,z_2)^{T}$ and
\begin{eqnarray}
z_1&=&\sin k_{x}+i\sin k_{y},\nonumber\\
z_2&=&\sin\lambda+i(\cos k_{x}+\cos k_{y}+\cos \lambda-m_{0}).
\end{eqnarray}
For general $m_{0}$, the explicit form of these $d_i$'s read
\begin{eqnarray}
d_{x}&=&2\sin k_{x}\sin\lambda+2\sin k_{y}(\cos k_{x}+\cos k_{y}+\cos\lambda-m_{0}),\nonumber\\
d_{y}&=&-2\sin k_{y}\sin\lambda+2\sin k_{x}(\cos k_{x}+\cos k_{y}+\cos\lambda-m_{0}),\nonumber\\
d_{z}&=&\sin^{2}k_{x}+\sin^{2}k_{y}-\sin^{2}\lambda-(\cos k_{x}+\cos k_{y}+\cos\lambda-m_{0})^{2}.\quad
\end{eqnarray}
For the special case $m_{0}=\frac{3}{2}$, they become
\begin{eqnarray}
d_{x}&=&2\sin k_{x}\sin\lambda+2\sin k_{y}\cos k_{x}+\sin 2k_{y}+2\sin k_{y}\cos\lambda\nn\\ &&-3\sin k_{y},\nonumber\\
d_{y}&=&-2\sin k_{y}\sin\lambda+2\sin k_{x}\cos k_{y}+\sin 2k_{x}+2\sin k_{x}\cos\lambda\nn\\ &&-3\sin k_{x},\nonumber\\
d_{z}&=&-\frac{13}{4}-\cos 2k_{x}-\cos 2k_{y}-2\cos k_{x}\cos k_{y}-2\cos k_{x}\cos\lambda\nonumber\\
&&-2\cos k_{y}\cos\lambda+3(\cos k_{x}+\cos k_{y}+\cos \lambda).
\end{eqnarray}

\section{II. Explicit expressions of BdG Hamiltonian in the real space: uniform system}

In the main article, the Hamiltonian is written in a compact form in the $\bk$ space. More explicit (less compact) expressions for the kinetic energy $\xi_\bk$ and the pairing gap $\Delta_\bk$ are \begin{eqnarray}
\xi_{\bk} &=& -\cos 2k_{x}-\cos 2k_{y}-2\cos k_{x}\cos k_{y}-2\cos k_{x}\cos\lambda\nonumber\\
&&-2\cos k_{y}\cos\lambda+3(\cos k_{x}+\cos k_{y}+\cos \lambda)-\frac{13}{4},\nonumber\\
\Delta_{\bk} 
&=& -2i\eta(\sin k_x +i\sin k_y)\nn\\ &&\times[e^{i\lambda} + (\cos k_x+\cos k_y -\frac{3}{2})].
\end{eqnarray}
For many purposes, it is useful to do a Fourier transformation to the real space.
For a uniform system (i.e. spatially independent $\lambda$) , the BdG Hamiltonian is given by $\hat{H}=\hat{H}_{0}+\hat{H}_{\Delta}$ with the kinetic energy term
\begin{widetext}
\begin{eqnarray}
\hat{H}_{0}&=&-\frac{1}{2} \sum_{x,y}\left\{\left[(2\cos\lambda-3)\left(c^{\dag}_{x,y}c_{x+1,y} +c^{\dag}_{x,y}c_{x,y+1}\right)\right.\right.
+\left(c^{\dag}_{x,y}c_{x+2,y}+c^{\dag}_{x,y}c_{x,y+2}\right)+\nonumber\\
&&\left.\left.\left(c^{\dag}_{x,y}c_{x+1,y+1} +c^{\dag}_{x,y}c_{x+1,y-1}\right)\right]+h.c.\right\}
-\sum_{x,y}(\frac{13}{4}-3\cos\lambda)c^{\dag}_{x,y}c_{x,y},\label{real-space-1} \eea
and the pairing term
\bea
\hat{H}_{\Delta}&=&\eta\sum_{x,y} \left\{\left[(\frac{3}{2}-\cos\lambda-i\sin\lambda)c^{\dag}_{x,y}c^{\dag}_{x+1,y}+h.c.\right]
+\left[(\sin\lambda-i\cos\lambda+\frac{3}{2}i)c^{\dag}_{x,y}c^{\dag}_{x,y+1}+h.c.\right] \right.\nonumber\\
&&\left.-\frac{1}{2}\left[c^{\dag}_{x,y}c^{\dag}_{x+2,y}+ic^{\dag}_{x,y}c^{\dag}_{x,y+2}+h.c.\right]
-\left[\frac{1+i}{2}c^{\dag}_{x,y}c^{\dag}_{x+1,y+1} +\frac{1-i}{2}c^{\dag}_{x,y}c^{\dag}_{x+1,y-1}+h.c.\right]\right\}, \label{real-space-2}
\end{eqnarray}
\end{widetext} where $x,y$ are real-space coordinates taking integer values. As explained in the main article, $\lambda$ is simply a parameter of the Hamiltonian. We can see that the pairing between the nearest-neighbor sites $(x,y)$ and $(x+1,y)$ is \bea \Delta_{(x,y);(x+1,y)}=\eta(\frac{3}{2}-\cos\lambda-i\sin\lambda),\eea and the pairing between sites $(x,y)$ and $(x,y+1)$ is \bea \Delta_{(x,y);(x,y+1)}=\eta(\sin\lambda -i\cos\lambda +\frac{3}{2}i), \eea thus \bea \Delta_{(x,y);(x,y+1)}=i\Delta_{(x,y);(x+1,y)}, \eea showing a $p$-wave character in real space. Since the Chern number is zero (for any value of parameter $\lambda$), the superconductor is topologically trivial, as we have explained in the main article.

\section{III. Explicit expressions of BdG Hamiltonian in the real space: with a defect}

In the previous section, we have focused on uniform systems with spatially independent $\lambda$. As explained in the main article, a topological defect can be created if the Hamiltonian parameter $\lambda$ depends on the polar angle $\theta$, which is measured from the defect center $(x_0,y_0)$, i.e. \begin{eqnarray}
 \theta_{x,y}= \arctan\frac{y-y_0}{x-x_0},
\end{eqnarray} such that the configuration cannot be smoothly deformed to a uniform one. The spatial dependence of the parameter $\lambda$ in our topological defects is given by \bea \lambda_{x,y}=n\theta_{x,y}.\eea
Before proceeding, it is useful to point out that, since the hopping terms and the pairing terms are defined on the links instead of sites, it is natural to take the middle point of the link to define $\theta$, for instance, $\cos\theta\, c^\dag_{x,y}c_{x+1,y}$ is taken as $\cos\theta_{x+1/2,y}\, c^\dag_{x,y}c_{x+1,y}$ with $\theta_{x+1/2,y}=\arctan[(y-y_0)/(x+1/2-x_0)]$. It is also viable to take a different convention, say $\cos\theta_{x,y}\, c^\dag_{x,y}c_{x+1,y}$, which does not affect the existence of Majorana zero mode, as verified in our numerical calculations. The reason is that $|\theta_{x,y}-\theta_{x+1/2,y}|\rw 0$ as $(x-x_0)^2+(y-y_0)^2\rw\infty$, thus different choices merely differ in the near-defect-core region. The topological character of the defect is fixed by the Hamiltonian far from the defect center, which is not affected by modifying the near-defect-core region.

With the above technical aspect explained, we are ready to write down the real-space BdG Hamiltonian in the presence of a defect:
\begin{widetext}
\begin{eqnarray}
\hat{H}_{0}&=&-\frac{1}{2}\sum_{xy}\left\{\left[(2\cos n\theta_{x+\frac{1}{2},y}-3)c^{\dag}_{x,y}c_{x+1,y}
+(2\cos n\theta_{x,y+\frac{1}{2}}-3)c^{\dag}_{x,y}c_{x,y+1}\right.\right.
+\left(c^{\dag}_{x,y}c_{x+2,y}+c^{\dag}_{x,y}c_{x,y+2}\right)+\nonumber\\
&&\left.\left.\left(c^{\dag}_{x,y}c_{x+1,y+1}+c^{\dag}_{x,y}c_{x+1,y-1}\right)\right]+h.c.\right\}
-\sum_{xy}(\frac{13}{4}-3\cos n\theta_{x,y}) c^{\dag}_{x,y}c_{x,y},\nonumber\\
\hat{H}_{\Delta}&=&\eta\sum_{xy}\left\{\left[(\frac{3}{2}-\cos n\theta_{x+\frac{1}{2},y}
-i\sin n\theta_{x+\frac{1}{2},y}) c^{\dag}_{x,y}c^{\dag}_{x+1,y}+h.c.\right]\right.\nonumber\\
&&+\left[(\sin n\theta_{x,y+\frac{1}{2}} -i\cos n\theta_{x,y+\frac{1}{2}}
+\frac{3}{2}i)c^{\dag}_{x,y}c^{\dag}_{x,y+1}+h.c.\right]\nonumber\\
&&\left.-\frac{1}{2}\left[c^{\dag}_{x,y}c^{\dag}_{x+2,y} +ic^{\dag}_{x,y}c^{\dag}_{x,y+2}+h.c.\right]
-\left[\frac{1+i}{2}c^{\dag}_{x,y}c^{\dag}_{x+1,y+1} +\frac{1-i}{2}c^{\dag}_{x,y}c^{\dag}_{x+1,y-1}+h.c.\right]\right\},
\end{eqnarray}
\end{widetext} which simply replace the parameter $\lambda$ in Eq.(\ref{real-space-1}) and Eq.(\ref{real-space-2}) by $n\theta$.

It is a conventional step to define $\Psi(x,y)=(c_{x,y},c_{x,y}^{\dag})^{T}$, and the real-space BdG Hamiltonian becomes \bea \hat{H}=\sum_{x,y;x',y'}\Psi^{\dag}(x,y)H_{x,y;x',y'}\Psi(x',y'), \eea with nonzero elements of $H_{x,y;x',y'}$ given by
\begin{widetext}
\begin{eqnarray}
H_{x,y;x,y}&=&\left(
                                             \begin{array}{cc}
                                               -\frac{13}{4}+3\cos n\theta_{x,y}  & 0 \\
                                               0 & \frac{13}{4}-3\cos n\theta_{x,y} \\
                                             \end{array}
                                           \right),\nonumber\\
H_{x,y;x\pm1,y}&=&\left(
                                             \begin{array}{cc}
                                               \frac{3}{2}-\cos n\theta_{x\pm\frac{1}{2},y}  & \mp i\sin n\theta_{x\pm\frac{1}{2},y}\mp\cos n\theta_{x\pm\frac{1}{2},y}\pm\frac{3}{2} \\
                                               \mp i\sin n\theta_{x\pm\frac{1}{2},y}\pm\cos n\theta_{x\pm\frac{1}{2},y}\mp\frac{3}{2} & -\frac{3}{2}+\cos n\theta_{x\pm\frac{1}{2},y} \\
                                             \end{array}
                                           \right),\nonumber\\
H_{x,y;x,y\pm1}&=&\left(
                                             \begin{array}{cc}
                                               \frac{3}{2}-\cos n\theta_{x,y\pm\frac{1}{2}}  & \mp i\cos n\theta_{x,y \pm\frac{1}{2}}\pm\frac{3}{2}i\pm\sin n\theta_{x,y\pm\frac{1}{2}} \\
                                               \mp i\cos n\theta_{x,y\pm\frac{1}{2}}\pm\frac{3}{2}i\mp\sin n\theta_{x,y\pm\frac{1}{2}} & -\frac{3}{2}+\cos n\theta_{x,y\pm\frac{1}{2}} \\
                                             \end{array}
                                           \right),\nonumber\\
H_{x,y;x\pm2,y}&=&\left(
                                             \begin{array}{cc}
                                               -\frac{1}{2}  & \mp\frac{1}{2}  \\
                                               \pm\frac{1}{2} & \frac{1}{2} \\
                                             \end{array}
                                           \right),\quad
H_{x,y;x,y\pm2}=\left(
                                             \begin{array}{cc}
                                               -\frac{1}{2}  & \mp\frac{1}{2}i  \\
                                               \mp\frac{1}{2}i & \frac{1}{2} \\
                                             \end{array}
                                           \right),\nonumber\\
H_{x,y;x\pm1,y\pm1}&=&\left(
                                             \begin{array}{cc}
                                               -\frac{1}{2}  & \mp\frac{1}{2}i\mp\frac{1}{2}  \\
                                               \mp\frac{1}{2}i\pm\frac{1}{2} & \frac{1}{2} \\
                                             \end{array}
                                           \right),\quad
                                           H_{x,y;x\pm1,y\mp1}=\left(
                                             \begin{array}{cc}
                                               -\frac{1}{2}  & \pm\frac{1}{2}i\mp\frac{1}{2}  \\
                                               \pm\frac{1}{2}i\pm\frac{1}{2} & \frac{1}{2} \\
                                             \end{array}
                                           \right).
\end{eqnarray}
\end{widetext}
The rank of the matrix $H_{x,y;x',y'}$ is $2L^2$, $L$ being the linear size of a square sample.

\section{IV. Numerical calculation of Hopf invariant}

We start from the $\bk$-space BdG Hamiltonian:
\bea H_\eta(\bk,\theta)\equiv \sum_{i}\tilde{d}_{i}\tau_{i}=\eta(d_{x}\tau_{x}+d_{y}\tau_{y})+d_{z}\tau_{z}. \eea
For $\eta\ll 1$, $H_\eta$ describes a superconductor with weak pairing (in the main article, we focused on the $\eta=1$ case).

\begin{figure}
\includegraphics[width=7cm, height=5cm]{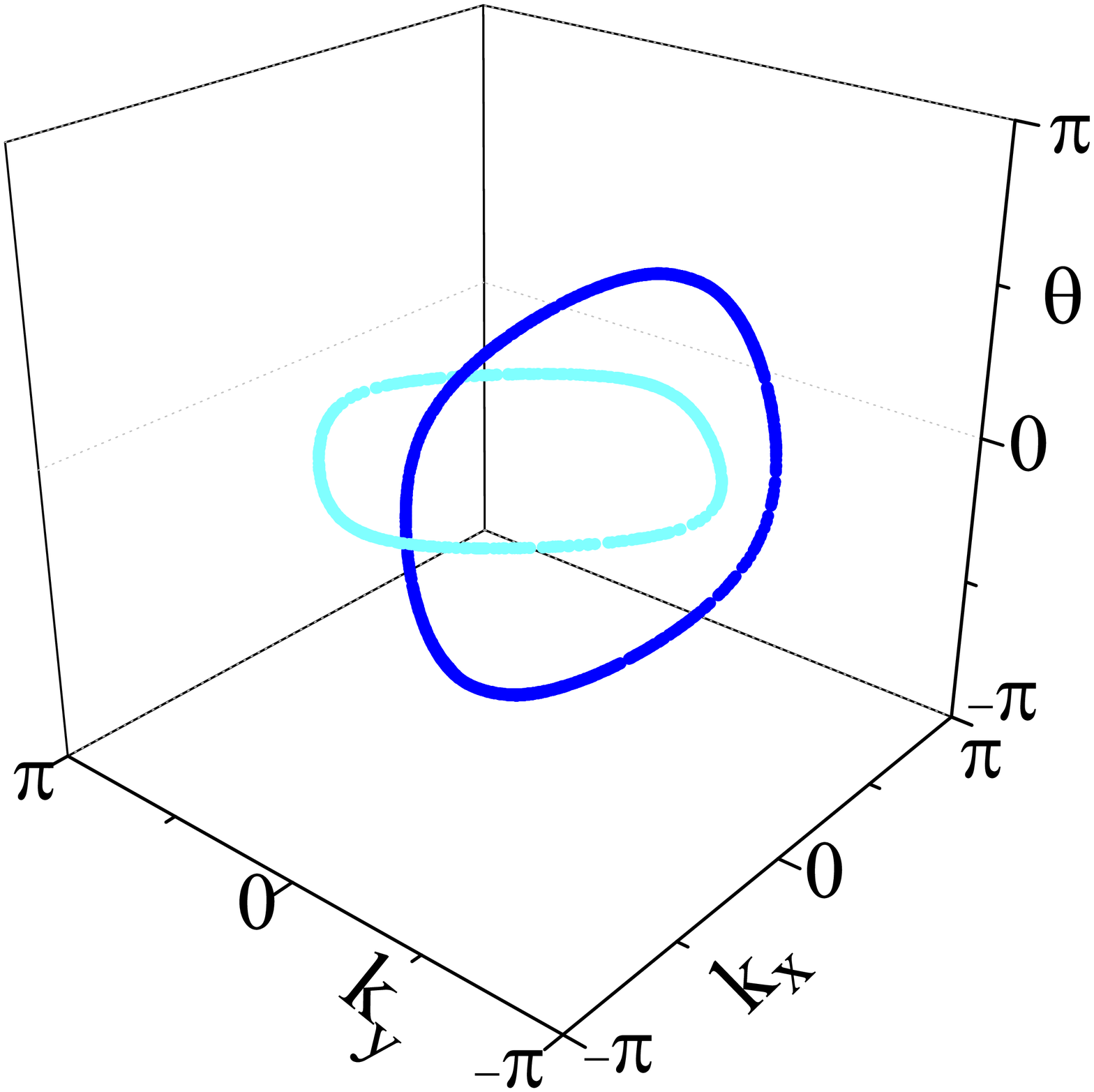}
\caption{The inverse images of $(0,0,1)$ (light cyan) and $(1,0,0)$ (dark blue) of the mapping ${\bf \hat{d}}(\bk,\theta): T^3\rw S^2$.
}  \label{linking}
\end{figure}

As we explained in the main article, the unit vector \bea {\bf \hat{d}}(\bk,\theta)\equiv \frac{1}{\sqrt{\eta^2(d_x^2+d_y^2)+d_z^2}} (\eta d_x,\eta d_y,d_z)\eea maps the 3D torus $T^3$ ($k_x,k_y,\theta$ are defined modulo $2\pi$) to the two-dimensional unit sphere $S^2$. For nonzero $n$, the inverse-image circles of two points on $S^2$ are linked, which is illustrated in Fig.\ref{linking} for the $n=1$ case.

The Hopf invariant, which characterizes the topological property of the Hamiltonian,
is given by
\begin{eqnarray}
N_{h}=-\int d^2kd\theta\,\bj(\bk,\theta)\cdot \bA(\bk,\theta),\label{integrationhopf}
\end{eqnarray}
where $\bj(\bk,\theta)=(j^x,j^y,j^\theta)$ (Note that the superscripts $x,y$ stand for $k_x,k_y$) takes the form of
\begin{eqnarray}
j^{\mu} = \frac{1}{8\pi}\epsilon^{\mu\nu\rho} \hat{{\bf d}}\cdot( \partial_\nu\hat{{\bf d}}
\times\partial_\rho\hat{{\bf d}})
\end{eqnarray}
and $\bA=(A^x,A^y,A^\theta)$ (again, the superscripts $x,y$ stand for $k_x,k_y$) is the gauge potential satisfying $\nabla \times \bA = \bj$, where $\nabla=(\partial/\partial k_x, \partial/\partial k_y, \partial/\partial\theta)$.
It is difficult to do the integration in Eq.(\ref{integrationhopf}) analytically. To do it numerically,
we rewrite Eq.(\ref{integrationhopf}) into a discrete form:
\begin{eqnarray}
N_{h}=-\frac{(2\pi)^{3}}{N^{3}}\sum_{\bk,\theta} \bj(\bk,\theta)\cdot \bA(\bk,\theta),\label{discretehopf}
\end{eqnarray}
where $N$ is the number of lattice sites (the lattice constant is set to unit)
and $k_{x,y},\theta$ take discrete values in
$\{-\pi,-\pi+\frac{2\pi}{N},...,\pi-\frac{2\pi}{N}\}$.

To obtain the expression of $\bA$ from $\bj$, we do the following Fourier transformation:
\begin{eqnarray}
\bA(\bk,\theta)&=&\frac{1}{N^{3/2}}\sum_{\bq}\bA(\bq)e^{-i(q_x k_x+q_yk_y+q_\theta\theta)},\nn\\
\bj(\bk,\theta)&=&\frac{1}{N^{3/2}}\sum_{\bq}\bj(\bq)e^{-i(q_xk_x+q_yk_y+q_\theta\theta)},
\end{eqnarray}
where $\bq=(q_x,q_y,q_\theta)$, whose components $q_{x,y,\theta}$ take discrete values
in $\{-\frac{N}{2},-\frac{N}{2}+1,...,\frac{N}{2}-1\}$.
Under the gauge $\bq\cdot\bA=0$, it is readily found that
\begin{eqnarray}
\bA(\bq)=-i\frac{\bq\times \bj(\bq)}{\bq^{2}},
\end{eqnarray}
thus the Hopf invariant becomes
\begin{eqnarray}
N_{h}&=&-\frac{(2\pi)^{3}}{N^{3}}\sum_{\bq} \bj(-\bq)\cdot \bA(\bq)\nn\\
&=&i\frac{(2\pi)^{3}}{N^{3}}\sum_{\bq}\frac{\bj(-\bq)\cdot (\bq\times \bj(\bq))}{\bq^{2}}.
\end{eqnarray}
The numerical integration converges quite rapidly as we increase $N$. For $\lambda=n\theta$, we find $N_{h}=n$. The $n=1$ case is shown in Fig.1b of the main article.

\section{V. More supporting data on the effects of impurity potential}

\begin{figure}
\subfigure{\includegraphics[width=7cm, height=5.5cm]{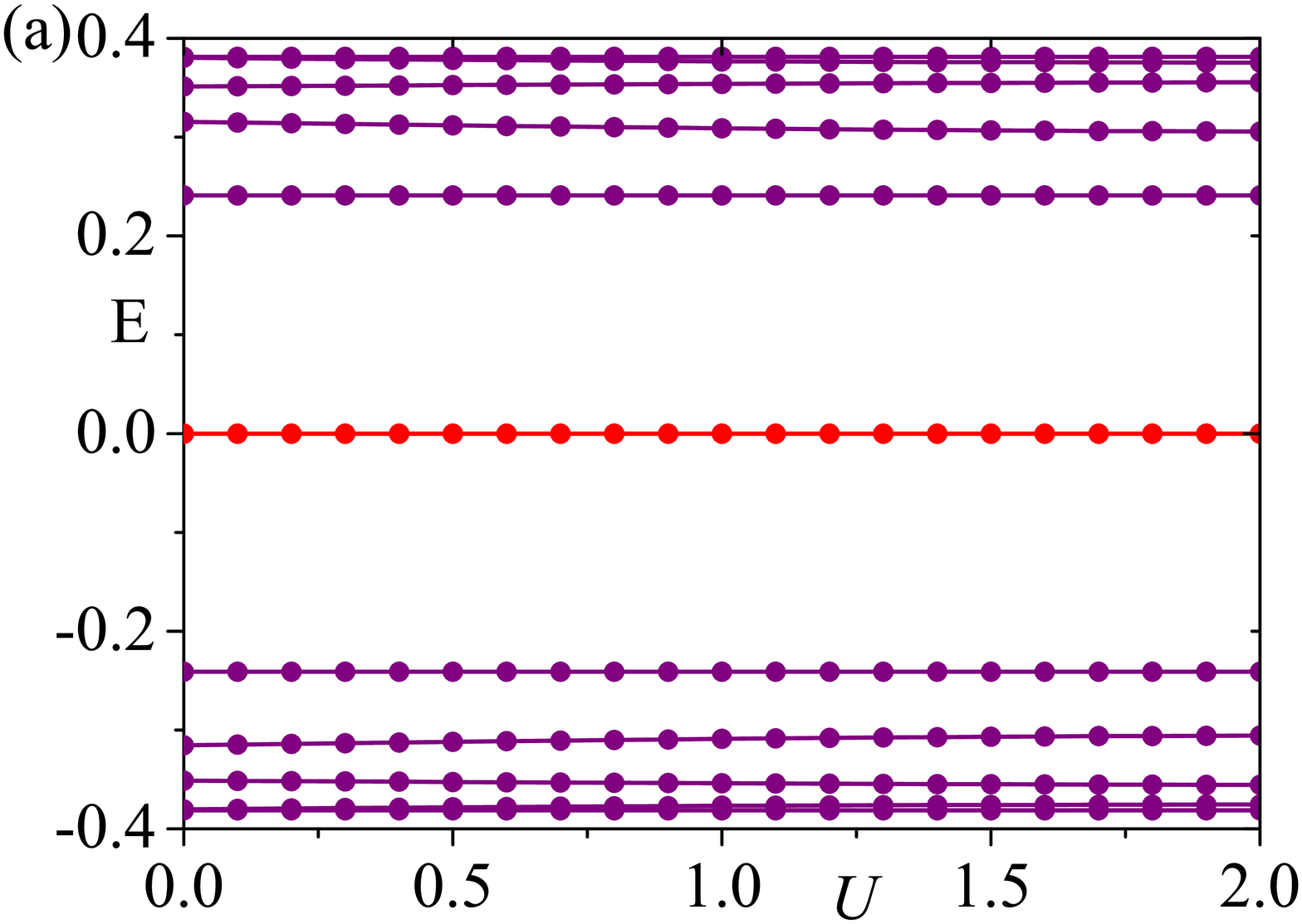}}
\subfigure{\includegraphics[width=7cm, height=5.5cm]{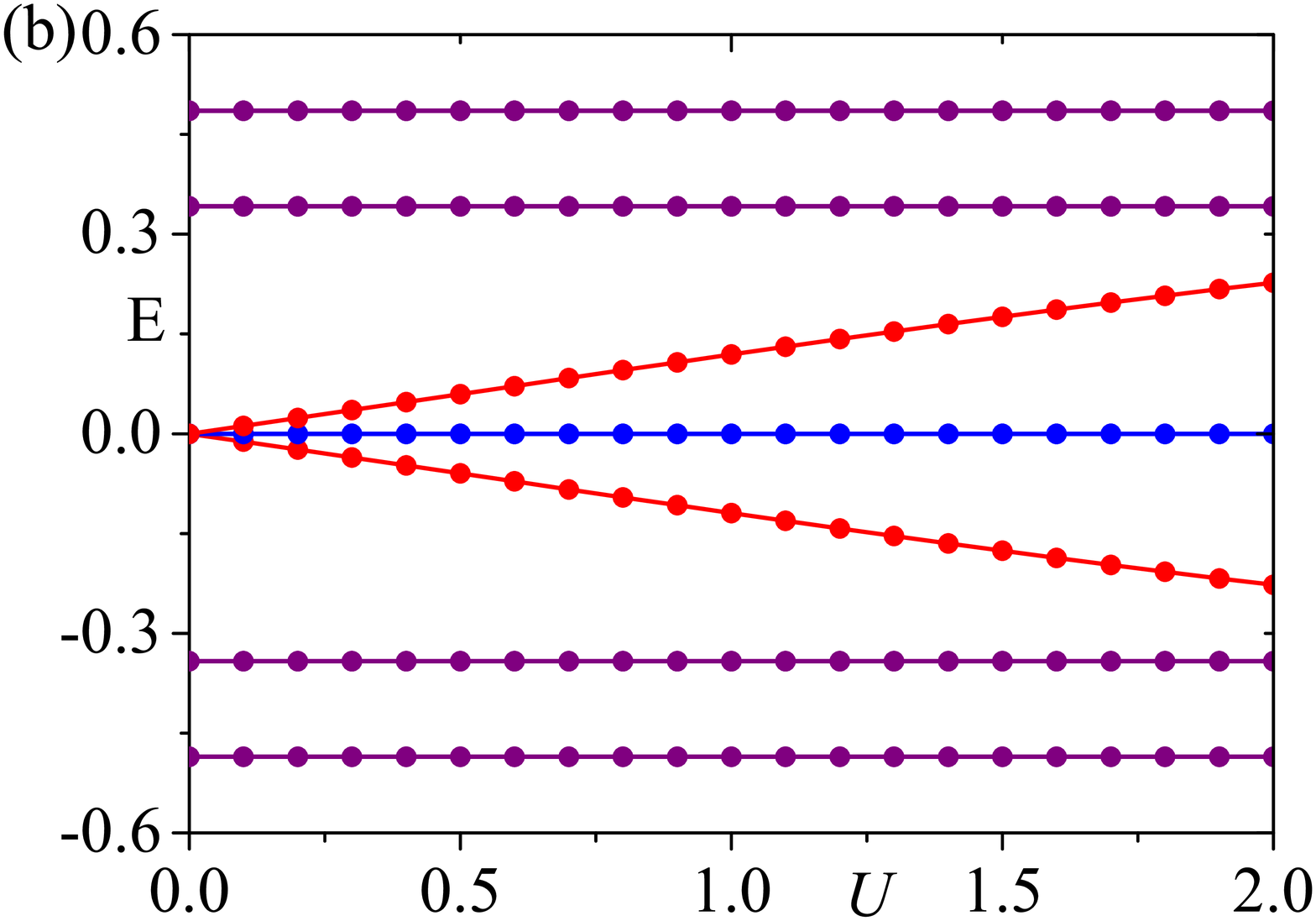}}
\subfigure{\includegraphics[width=7cm, height=5.5cm]{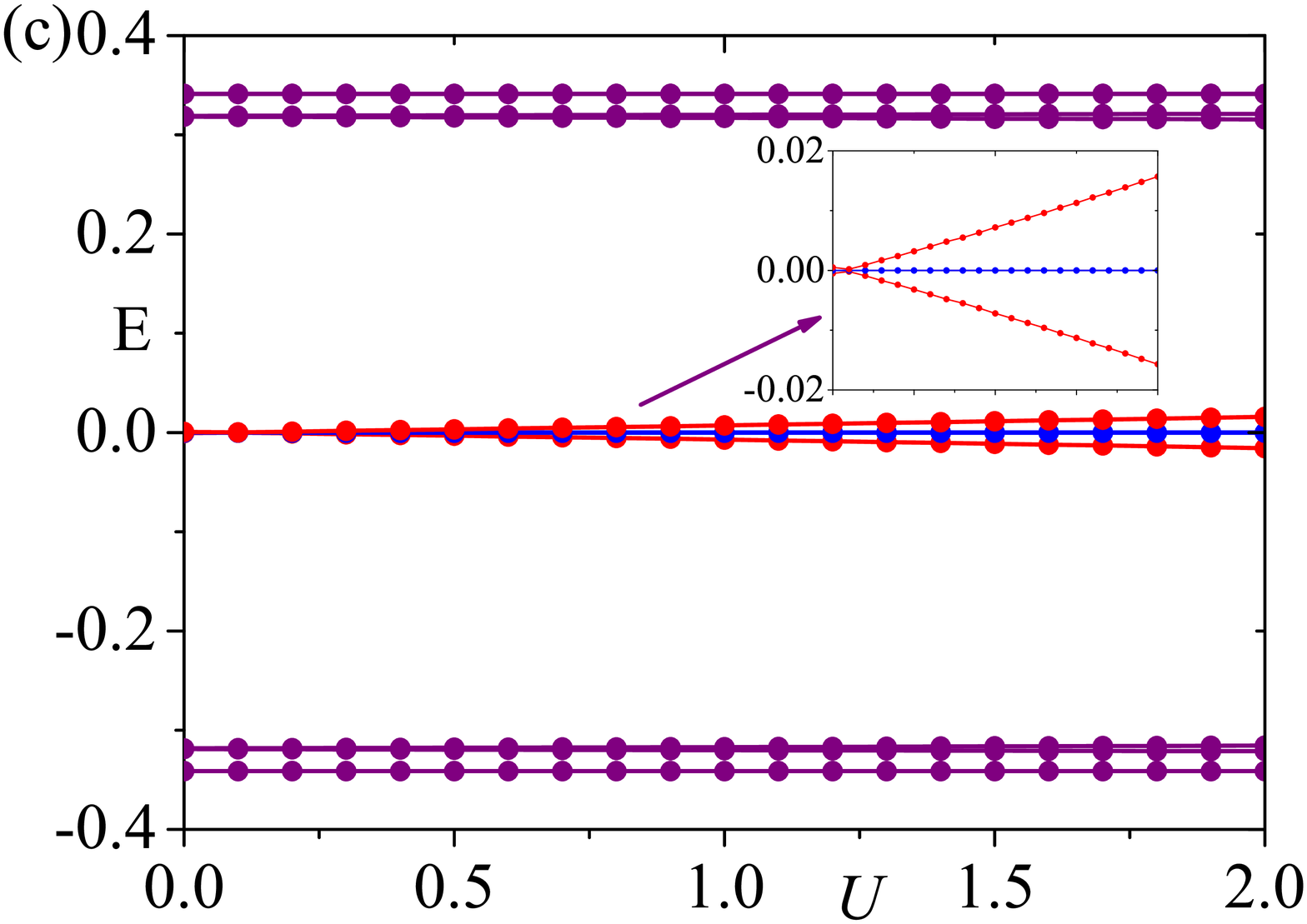}}
\caption{ The energy eigenvalues close to $E=0$ for three defects, as a function of an impurity potential $U$ at a single site $A$ (specified in Fig.1a in the main article). (a) $m_{0}=3/2$ and $n=1$, with Hopf invariant $N_{h}=1$.
(b) $m_{0}=0$ and $n=1$, with $N_{h}=-2$.  (c) $m_{0}=3/2$, and $n=2$, with $N_{h}=2$. The inset of (c) provides
a zoom-in view of the energy splitting. Several nonzero energy eigenvalues are almost independent of $U$ in (a), (b), and (c).  The reason is that their eigenfunctions are quite far from the impurity site $A$.  }  \label{Z2}
\end{figure}

In the main article, we have shown that the Majorana zero mode for $n=1$ is robust to an impurity potential, and remarked that robust Majorana zero mode is absent for even-integer Hopf invariant. Here, more data on this is shown in Fig.\ref{Z2}. As explained in the main article, an impurity potential $Uc^\dag_A c_A$ at a single site $A$ (indicated in Fig.1a in the main article) is added to test the robustness of Majorana zero modes. For odd-integer Hopf invariant, Fig.\ref{Z2}a shows that the impurity potential cannot move the zero mode energies away from $E=0$ (in consistent with the results given in the main article). For even-integer Hopf invariant, as shown in Fig.\ref{Z2}b and Fig.\ref{Z2}c, zero modes are generally absent in the defect in the presence of impurity potential (the blue lines in Fig.\ref{Z2}b and Fig.\ref{Z2}c are the energies for modes near the boundary, not near the defect at the system center). Note that several nonzero energy eigenvalues are almost independent of $U$ in Fig.\ref{Z2}a, Fig.\ref{Z2}b, and Fig.\ref{Z2}c, because their eigenfunctions are quite far from the impurity site $A$.

\section{VI. Edge theory for various sample geometries}

\begin{figure}
\includegraphics[width=8cm, height=6cm]{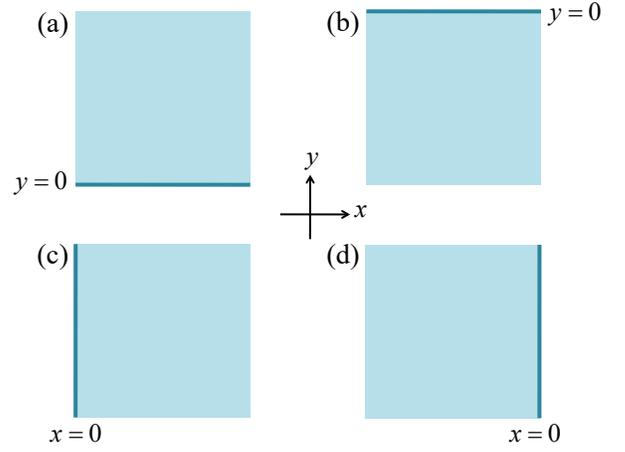}
\caption{Four geometries of the sample. The sample occupies
(a) the $y>0$ region, (b) the $y<0$ region, (c) the $x>0$ region,
(d) the $x<0$ region.}  \label{boundary}
\end{figure}

\subsection{Sample occupying $y>0$}

First, we study a semi-infinite geometry with the sample occupying the $y>0$ region (illustrated
in Fig.\ref{boundary}a). The momentum $k_x$ remains a good quantum number, while $k_y$ is not. In the $y$ direction, we use the real-space coordinate, which is an integer in our lattice model.     The wave functions and energy eigenvalues can be obtained by solving the following equation
\begin{eqnarray}
\left(
  \begin{array}{ccccccc}
    H_{0} & H_{1} & H_{2} & 0& 0 & 0 & \cdots\\
    H_{1}^{\dag} & H_{0} & H_{1} & H_{2} & 0 & 0 & \cdots \\
    H_{2}^{\dag} & H_{1}^{\dag} & H_{0} & H_{1} & H_{2} & 0 & \cdots \\
    0 & H_{2}^{\dag} & H_{1}^{\dag} & H_{0} & H_{1} & H_{2} & \cdots \\
    \vdots & \vdots & \vdots & \vdots & \vdots & \vdots & \ddots \\
  \end{array}
\right)\left(
         \begin{array}{c}
           \psi_{1} \\
           \psi_{2} \\
           \psi_{3} \\
           \psi_{4} \\
           \vdots \\
         \end{array}
       \right)=E\left(
         \begin{array}{c}
           \psi_{1} \\
           \psi_{2} \\
           \psi_{3} \\
           \psi_{4} \\
           \vdots \\
         \end{array}
       \right),\label{y}
\end{eqnarray}
where the subscript $j$ of $\psi_j$ is the $y$ coordinate, each $\psi_j$ is two-component, and
\begin{widetext}
\begin{eqnarray}
H_{0}(k_{x},\lambda)&=&\left(
        \begin{array}{cc}
          -\frac{13}{4}-\cos2k_{x}+3(\cos k_{x}+\cos\lambda)-2\cos k_{x}\cos\lambda & 2\sin k_{x}\sin\lambda-i(\sin2k_{x}+2\sin k_{x}\cos\lambda-3\sin k_{x}) \\
          2\sin k_{x}\sin\lambda+i(\sin2k_{x}+2\sin k_{x}\cos\lambda-3\sin k_{x}) & \frac{13}{4}+\cos2k_{x}-3(\cos k_{x}+\cos\lambda)+2\cos k_{x}\cos\lambda \\
        \end{array}
      \right),\nonumber\\
      H_{1}(k_{x},\lambda)&=&\left(
        \begin{array}{cc}
          \frac{3}{2}-\cos k_{x}-\cos\lambda & -i (\cos k_{x}+\cos\lambda-\frac{3}{2})-i\sin k_{x}+\sin\lambda \\
          -i (\cos k_{x}+\cos\lambda-\frac{3}{2})+i\sin k_{x}-\sin\lambda  & -\frac{3}{2}+\cos k_{x}+\cos\lambda\\
        \end{array}
      \right),\nonumber\\
H_{2}(k_{x},\lambda)&=&\left(
        \begin{array}{cc}
          -\frac{1}{2} & -\frac{1}{2}i  \\
         -\frac{1}{2}i  & \frac{1}{2}\\
        \end{array}
      \right).\nonumber\\
\end{eqnarray}   \end{widetext} Eq.(\ref{y}) can be written compactly as $H(k_x,\lambda)|\Psi\ra=E|\Psi\ra$, $H$ denoting the large matrix at the left-hand-side of Eq.(\ref{y}).
Let us first find the zero-energy ($E=0$) solutions, if any, at $k_{x}=0$ and $\lambda=0$. When $k_{x}=0$ and $\lambda=0$, $H_i$'s simplify to
\begin{eqnarray}
H_{0}(0,0)&=&\left(
        \begin{array}{cc}
          -\frac{1}{4} & 0 \\
          0 & \frac{1}{4} \\
        \end{array}
      \right),\, \nn\\
H_{1}(0,0)&=&H_2(0,0)=\left(
        \begin{array}{cc}
          -\frac{1}{2} & -\frac{1}{2}i  \\
         -\frac{1}{2}i  & \frac{1}{2}\\
        \end{array}
      \right).
\end{eqnarray}
We can define $\tau_\pm =(\tau_z\pm i\tau_x)/2$, which is the raising/lowering operator of $\tau_y$, thus $H_0=-(\tau_+ +\tau_-)/4$, $H_1=H_2=-\tau_+$, and $H_1^\dag=H_2^\dag=-\tau_-$. Operating on the $\tau_y$ eigenvectors $|\chi_{\pm}\rangle=(1,\pm i)^{T}/\sqrt{2}$, they produce
\begin{eqnarray}
H_{1}|\chi_{+}\rangle&=&0,\,
H_{1}^{\dag}|\chi_{+}\rangle=-|\chi_{-}\rangle; \nonumber\\
H_{1}|\chi_{-}\rangle&=&-|\chi_{+}\rangle,\,
H_{1}^{\dag}|\chi_{-}\rangle=0; \nonumber\\
H_{0}|\chi_{+}\rangle&=&-\frac{1}{4}|\chi_{-}\rangle, \,H_{0}|\chi_{-}\rangle=-\frac{1}{4}|\chi_{+}\rangle,
\end{eqnarray}
We can check that the zero-energy wavefunctions take the form of \bea |\Psi\rangle=\sum_{j}a_{j}|j\rangle\otimes|\chi_{-}\rangle, \eea where $|j\rangle$ is localized on the $j$-site:
\begin{eqnarray}
|j\rangle=(\underbrace{0,0,...,0}_{j-1},1,0,...)^{T},
\end{eqnarray}
with the coefficients $a_{j}$ satisfying the following iteration relation:
\begin{eqnarray}
a_{j}=-4(a_{j+1}+a_{j+2}).
\end{eqnarray} We can see that $\psi_j=a_j|\chi_-\ra$.
It is straightforward to find two normalizable solutions for $a_{j}$'s, which we will denote as $a_j=\alpha_{j}$ ($j=1,2,3,\cdots$) and $a_j=\beta_{j}$ ($j=1,2,3,\cdots$). The explicit expressions of $\alpha_j$ and $\beta_j$ are \bea \alpha_{j}=C_{\alpha}\frac{(-1)^{j+1}}{2^{j-1}}, \eea and \bea \beta_{j}=C_{\beta}\frac{(-1)^{j}(j-1)}{2^{j-2}},\eea where
$C_{\alpha}$ and $C_{\beta}$ are two normalization constants such that $\sum_{j}|\alpha_{j}|^{2}=1$ and $\sum_{j}|\beta_{j}|^{2}=1$. It is straightforward to find that $C_{\alpha}=\sqrt{3}/2$ and $C_{\beta}=\sqrt{27/80}$.

We can check that the two wave functions  $|\Psi_{1}\rangle=\sum_{j}\alpha_{j}|j\rangle\otimes|\chi_{-}\rangle$
and $|\Psi_{2}\rangle=\sum_{j}\beta_{j}|j\rangle\otimes|\chi_{-}\rangle$ are not orthogonal. The orthogonalization
can be achieved by the Gram-Schmidt orthogonalization. We have
\begin{eqnarray}
|\Psi_{1}^{\rm o}\rangle&=&|\Psi_{1}\rangle,\nonumber\\
|\Psi_{2}^{\rm o}\rangle&=&\mathcal{N}(|\Psi_{2}\rangle- |\Psi_{1} \rangle\langle\Psi_{1}|\Psi_{2}\rangle),
\end{eqnarray}
where $\mathcal{N}=\sqrt{5}/2$ is a normalization constant
and the superscript ``o'' stands for
``orthogonalization''.

So far, we have focused on $k_{x}=0$, $\lambda=0$.
In the neighbourhood of $k_{x}=0$, $\lambda=0$, we can expand the Hamiltonian $H$ to the first order of $k_{x}$ and $\lambda$, namely,
$H_{i}(k_{x},\lambda)=H_{i}(0,0)+\Delta H_{i}(k_{x},\lambda)$ with
\begin{eqnarray}
\Delta H_{0}(k_{x},\lambda)&=&\left(
        \begin{array}{cc}
          0 &  -ik_{x} \\
          ik_{x} & 0 \\
        \end{array}
      \right),\, \nn\\
\Delta H_{1}(k_{x},\lambda)&=&\left(
        \begin{array}{cc}
          0 & \lambda-ik_{x} \\
          -\lambda+ik_{x} & 0 \\
        \end{array}
      \right).
\end{eqnarray}
It is readily seen that
\begin{eqnarray}
\langle\chi_{-}|\Delta H_{0}|\chi_{-}\rangle&=&-k_{x}, \,\langle\chi_{-}|\Delta H_{1}|\chi_{-}\rangle=-(k_{x}+i\lambda),\,\nn\\
\langle\chi_{-}|\Delta H_{1}^{\dag}|\chi_{-}\rangle&=&-(k_{x}-i\lambda).
\end{eqnarray}
Straightforward calculations lead to
\bea
\langle\Psi_{1}^{\rm o}|\Delta H| \Psi_{1}^{\rm o}\rangle&=&-\left(\sum_{j=1}\alpha_{j}^{2}+2\sum_{j=1}\alpha_{j}\alpha_{j+1}\right)k_{x}\nn\\
&=&-C_{\alpha}^{2}\left(\sum_{j=1}\frac{1}{2^{2j-2}}-2\sum_{j=1}\frac{1}{2^{2j-1}}\right)k_{x}\nn \\&=&0,\nn \eea
and \bea
\langle\Psi_{2}^{\rm o}|\Delta H| \Psi_{2}^{\rm o}\rangle&=&-\mathcal{N}^{2}\left[\sum_{j}[\beta_{j}
-(\sum_{l}\alpha_{l}\beta_{l})\alpha_{j}]^{2}+2\sum_{j}\right.\nn\\
&&\left.[\beta_{j}-(\sum_{l}\alpha_{l}\beta_{l}) \alpha_{j}][\beta_{j+1}-(\sum_{l}\alpha_{l}\beta_{l})\alpha_{j+1}]\right]k_{x}\nonumber\\
&=&-\mathcal{N}^{2}\left[(\sum_{l}\alpha_{l}\beta_{l})^{2}\left(\sum_{j=1}\alpha_{j}^{2}+2\sum_{j=1}\alpha_{j}\alpha_{j+1}\right)\right.\nn\\
&&+\left(\sum_{j=1}\beta_{j}^{2}+2\sum_{j=1}\beta_{j}\beta_{j+1}\right)-2(\sum_{l}\alpha_{l}\beta_{l})\nn\\
&&\left.\times\sum_{j}(\alpha_{j}\beta_{j}+\alpha_{j}\beta_{j+1}+\alpha_{j+1}\beta_{j})\right]k_{x}\nn\\
&&=-\mathcal{N}^{2}\left[\left(\sum_{j=1}\beta_{j}^{2}+2\sum_{j=1}\beta_{j}\beta_{j+1}\right)-2(\sum_{l}\alpha_{l}\beta_{l})\right.\nn\\
&&\left.\times\sum_{j}(\alpha_{j}\beta_{j}+\alpha_{j}\beta_{j+1}+\alpha_{j+1}\beta_{j})\right]k_{x}\nn\\
&=&-\mathcal{N}^{2}\left[C_{\beta}^{2}\sum_{j=1} \frac{1-j}{2^{2j-4}}-2(\sum_{l}\alpha_{l}\beta_{l})C_{\alpha}C_{\beta}\sum_{j}\frac{1}{2^{2j-2}}\right]k_{x}.\nn
\eea
Noting the mathematical identities
\begin{eqnarray}
\sum_{j}\alpha_{j}\beta_{j}&=&C_{\alpha}C_{\beta}\sum_{j}\frac{1-j}{2^{2j-3}},\nn\\
\sum_{j}\frac{1}{2^{2j-2}}&=&\frac{1}{C_{\alpha}^{2}},
\end{eqnarray}
we can see that
\begin{eqnarray}
\langle\Psi_{2}^{\rm o}|\Delta H| \Psi_{2}^{\rm o}\rangle &=& -\mathcal{N}^{2}\left[C_{\beta}^{2}
\frac{2(\sum_{l}\alpha_{l}\beta_{l})}{C_{\alpha}C_{\beta}}-2(\sum_{l}\alpha_{l}\beta_{l})
C_{\alpha}C_{\beta}\frac{1}{C_\alpha^{2}}\right]k_{x}\nn\\
&=&0.
\end{eqnarray}
Similarly, we find that
 \begin{eqnarray}
\langle\Psi_{1}^{\rm o}|\Delta H| \Psi_{2}^{\rm o}\rangle&=&-\mathcal{N}
\left[\sum_{j}\left(\alpha_{j}[\beta_{j+1}-(\sum_{l}\alpha_{l}\beta_{l})\alpha_{i+1}]\right.\right.\nn\\
&&\left.+\alpha_{j+1}[\beta_{j}-(\sum_{l}\alpha_{l}\beta_{l})\alpha_{i}]\right)k_{x}\nn\\
&&+\sum_{j}\left(\alpha_{j}[\beta_{j+1}-(\sum_{l}\alpha_{l}\beta_{l})\alpha_{i+1}]\right.\nn\\
&&\left.\left.-\alpha_{j+1}[\beta_{j}-(\sum_{l}\alpha_{l}\beta_{l})\alpha_{i}]\right)i\lambda\right]\nn\\
&=&-\mathcal{N}\left[\left(\sum_{j}[\alpha_{j}\beta_{j+1}+\alpha_{j+1}\beta_{j}
-2(\sum_{l}\alpha_{l}\beta_{l})\alpha_{i}\alpha_{i+1}]\right)k_{x}\right.\nn\\
&&\left.+\sum_{j}(\alpha_{j}\beta_{j+1}-\alpha_{j+1}\beta_{j})i\lambda\right]\nn\\
&=&-\mathcal{N}\left(C_{\alpha}C_{\beta}\sum_{j}\frac{1}{2^{2j-2}}\right)(k_{x}+i\lambda)\nn\\
&=&-\frac{\mathcal{N}C_{\beta}}{C_{\alpha}}(k_{x}+i\lambda)\nn\\
&=&-\frac{3}{4}(k_{x}+i\lambda),\nn\\
\langle\Psi_{2}^{\rm o}|\Delta H| \Psi_{1}^{\rm o}\rangle &=&
\langle\Psi_{1}^{\rm o}|\Delta H| \Psi_{2}^{\rm o}\rangle^{*} \nn \\ &=&-\frac{3}{4}(k_{x}-i\lambda). \nn
\end{eqnarray}
To summarize the above calculations, we have the following low-energy effective Hamiltonian for the edge states
near $k_x=0$, $\lambda=0$:
\begin{eqnarray}
H_{\rm eff}&=&\left(
  \begin{array}{cc}
    \langle\Psi_{1}^{\rm o}|\Delta H| \Psi_{1}^{\rm o}\rangle & \langle\Psi_{1}^{\rm o}|\Delta H| \Psi_{2}^{\rm o}\rangle \\
    \langle\Psi_{2}^{\rm o}|\Delta H| \Psi_{1}^{\rm o}\rangle & \langle\Psi_{2}^{\rm o}|\Delta H| \Psi_{2}^{\rm o}\rangle \\
  \end{array}
\right) \nn \\ &=&-\frac{3}{4}\left(
  \begin{array}{cc}
    0 & k_{x}+i\lambda  \\
    k_{x}-i\lambda & 0 \\
  \end{array}
\right),
\end{eqnarray}
or more compactly,
\begin{eqnarray}
H_{\rm eff}=\frac{3}{4}(-k_{x}\sigma_{x}+\lambda \sigma_{y}).
\end{eqnarray}

\subsection{Sample occupying $y<0$}

Now we study the sample occupying the $y<0$ region (Fig.\ref{boundary}b), then the edge modes
can be obtained by solving the following eigenvalue problem,
\begin{eqnarray}
\left(
  \begin{array}{ccccccc}
    H_{0} & H_{1}^{\dag} & H_{2}^{\dag} & 0& 0 & 0 & \cdots\\
    H_{1} & H_{0} & H_{1}^{\dag} & H_{2}^{\dag} & 0 & 0 & \cdots \\
    H_{2} & H_{1} & H_{0} & H_{1}^{\dag} & H_{2}^{\dag} & 0 & \cdots \\
    0 & H_{2} & H_{1} & H_{0} & H_{1}^{\dag} & H_{2}^{\dag} & \cdots \\
    \vdots & \vdots & \vdots & \vdots & \vdots & \vdots & \ddots \\
  \end{array}
\right)\left(
         \begin{array}{c}
           \psi_{-1} \\
           \psi_{-2} \\
           \psi_{-3} \\
           \psi_{-4} \\
           \vdots \\
         \end{array}
       \right)=E\left(
         \begin{array}{c}
           \psi_{-1} \\
           \psi_{-2} \\
           \psi_{-3} \\
           \psi_{-4} \\
           \vdots \\
         \end{array}
       \right).
\end{eqnarray}
Following the same procedures in the previous section, we first find the solutions for $k_{x}=0$ and $\lambda=0$.
We find two wave functions with $E=0$, one of which is
$|\Psi_{1}'\rangle=\sum_{j}\alpha_{j}|j\rangle\otimes|\chi_{+}\rangle$,
and the other is $|\Psi_{2}'\rangle=\sum_{j}\beta_{j}|j\rangle\otimes|\chi_{+}\rangle$.
The orthogonalzation of the two wavefunctions are achieved by the
Gram-Schmidt orthogonalization,
\begin{eqnarray}
|\Psi_{1}^{'\rm o}\rangle&=&|\Psi_{1}'\rangle,\nonumber\\
|\Psi_{2}^{'\rm o}\rangle&=&\mathcal{N}(|\Psi_{2}'\rangle -|\Psi'_{1} \rangle\langle\Psi'_{1}|\Psi_{2}'\rangle).
\end{eqnarray}
It is readily checked that
\begin{eqnarray}
\langle\chi_{+}|\Delta H_{0}|\chi_{+}\rangle&=&k_{x}, \,\langle\chi_{+}|\Delta H_{1}|\chi_{+}\rangle=k_{x}+i\lambda,\,\nn\\
\langle\chi_{+}|\Delta H_{1}^{\dag}|\chi_{+}\rangle&=&k_{x}-i\lambda.
\end{eqnarray}
Straightforward calculations yield
\begin{eqnarray}
\left(
  \begin{array}{cc}
    \langle\Psi_{1}^{'\rm o}|\Delta H |\Psi_{1}^{'\rm o}\rangle & \langle\Psi_{1}^{'\rm o}|\Delta H |\Psi_{2}^{'\rm o}\rangle \\
    \langle\Psi_{2}^{'\rm o}|\Delta H |\Psi_{1}^{'\rm o}\rangle & \langle\Psi_{2}^{'\rm o}|\Delta H |\Psi_{2}^{'\rm o}\rangle \\
  \end{array}
\right)=\frac{3}{4}\left(
  \begin{array}{cc}
    0 & k_{x}-i\lambda   \\
    k_{x}+i\lambda   & 0 \\
  \end{array}
\right),
\end{eqnarray}
thus, the low-energy effective Hamiltonian for the edge state
reads
\begin{eqnarray}
H_{\rm eff}=\frac{3}{4}(k_{x}\sigma_{x}+\lambda \sigma_{y}).
\end{eqnarray}

\subsection{Sample occupying $x>0$}

For sample occupying the $x>0$ region (Fig.\ref{boundary}c),
the edge modes can be obtained by solving the following eigenvalue problem,
\begin{eqnarray}
\left(
  \begin{array}{ccccccc}
    \tilde{H}_{0} & \tilde{H}_{1} & \tilde{H}_{2} & 0& 0 & 0 & \cdots\\
    \tilde{H}_{1}^{\dag} & \tilde{H}_{0} & \tilde{H}_{1} & \tilde{H}_{2} & 0 & 0 & \cdots \\
    \tilde{H}_{2}^{\dag} & \tilde{H}_{1}^{\dag} & \tilde{H}_{0} & \tilde{H}_{1} & \tilde{H}_{2} & 0 & \cdots \\
    0 & \tilde{H}_{2}^{\dag} & \tilde{H}_{1}^{\dag} & \tilde{H}_{0} & \tilde{H}_{1} & \tilde{H}_{2} & \cdots \\
    \vdots & \vdots & \vdots & \vdots & \vdots & \vdots & \ddots \\
  \end{array}
\right)\left(
         \begin{array}{c}
           \tilde{\psi}_{1} \\
           \tilde{\psi}_{2} \\
           \tilde{\psi}_{3} \\
           \tilde{\psi}_{4} \\
           \vdots \\
         \end{array}
       \right)=E\left(
         \begin{array}{c}
           \tilde{\psi}_{1} \\
           \tilde{\psi}_{2} \\
           \tilde{\psi}_{3} \\
           \tilde{\psi}_{4} \\
           \vdots \\
         \end{array}
       \right),
\end{eqnarray}
where
\begin{widetext}
\begin{eqnarray}
\tilde{H}_{0}(k_{y},\lambda)&=&\left(
        \begin{array}{cc}
          -\frac{13}{4}-\cos2k_{y}+3(\cos k_{y}+\cos\lambda)-2\cos k_{y}\cos\lambda & 2i\sin k_{y}\sin\lambda+(\sin2k_{y}+2\sin k_{y}\cos\lambda-3\sin k_{y}) \\
          -2i\sin k_{y}\sin\lambda+(\sin2k_{y}+2\sin k_{y}\cos\lambda-3\sin k_{y}) & \frac{13}{4}+\cos2k_{y}-3(\cos k_{y}+\cos\lambda)+2\cos k_{y}\cos\lambda \\
        \end{array}
      \right),\nonumber\\
\tilde{H}_{1}(k_{y},\lambda)&=&\left(
        \begin{array}{cc}
          \frac{3}{2}-\cos k_{y}-\cos\lambda & -i\sin\lambda+\sin k_{y}-(\cos k_{y}+\cos\lambda-\frac{3}{2}) \\
          -i\sin\lambda+\sin k_{y}+(\cos k_{y}+\cos\lambda-\frac{3}{2})  & -\frac{3}{2}+\cos k_{y}+\cos\lambda\\
        \end{array}
      \right),\nonumber\\
\tilde{H}_{2}(k_{y},\lambda)&=&\left(
        \begin{array}{cc}
          -\frac{1}{2} & -\frac{1}{2} \\
         \frac{1}{2}  & \frac{1}{2}\\
        \end{array}
      \right).\nonumber\\
\end{eqnarray}
\end{widetext}
When $k_{y}=0$ and $\lambda=0$,
\begin{eqnarray}
\tilde{H}_{0}=\left(
        \begin{array}{cc}
          -\frac{1}{4} & 0 \\
          0 & \frac{1}{4} \\
        \end{array}
      \right),
\tilde{H}_{1}=\tilde{H}_2=\left(
        \begin{array}{cc}
          -\frac{1}{2} & -\frac{1}{2}  \\
         \frac{1}{2}  & \frac{1}{2}\\
        \end{array}
      \right).
\end{eqnarray} It is not difficult to see that
\begin{eqnarray}
\tilde{H}_{1}|\xi_{+}\rangle&=&0,\, \tilde{H}_{1}^{\dag}|\xi_{+}\rangle=-\xi_{-},\, \nonumber\\
\tilde{H}_{1}|\xi_{-}\rangle&=&-|\xi_{+}\rangle,\, \tilde{H}_{1}^{\dag}|\xi_{-}\rangle=0, \nonumber\\
\tilde{H}_{0}|\xi_{+}\rangle &=&-\frac{1}{4}|\xi_{-}\rangle,\,\tilde{H}_{0}|\xi_{-}\rangle=-\frac{1}{4}|\xi_{+}\rangle,
\end{eqnarray}
where $|\xi_\pm\ra=(1,\pm1)^T/\sqrt{2}$ are the two eigenvectors of $\tau_{x}$.
Following similar procedures as previous sections, we find
two solutions at $E=0$ for $k_y=\lambda=0$. One is
$|\tilde{\Psi}_{1}\rangle=\sum_{j}\alpha_{j}|j\rangle\otimes|\xi_{+}\rangle$,
and the other is $|\tilde{\Psi}_{2}\rangle=\sum_{j}\beta_{j}|j\rangle\otimes|\xi_{+}\rangle$.
Again we adopt the Gram-Schmidt orthogonalization to define
\begin{eqnarray}
|\tilde{\Psi}_{1}^{\rm o}\rangle&=&|\tilde{\Psi}_{1}\rangle,\nn\\
|\tilde{\Psi}_{2}^{\rm o}\rangle&=&\mathcal{N}(|\tilde{\Psi}_{2}\rangle- |\tilde{\Psi}_{1}\rangle\langle\tilde{\Psi}_{1}|\tilde{\Psi}_{2}\rangle).
\end{eqnarray}
In the neighbourhood of $k_{y}=0$ and $\lambda=0$, $\tilde{H}_{i=0,1}$ can be expanded to the first order of $k_{y}$ and $\lambda$:
\bea \tilde{H}_{i}(k_{y},\lambda)=\tilde{H}_{i}(0,0)+\Delta \tilde{H}_{i}(k_{y},\lambda), \eea with
\begin{eqnarray}
\Delta \tilde{H}_{0}=\left(
        \begin{array}{cc}
          0 & k_{y} \\
          k_{y} & 0 \\
        \end{array}
      \right),\,
\Delta \tilde{H}_{1}=\left(
        \begin{array}{cc}
          0 &  k_{y}-i\lambda \\
          k_{y}-i\lambda & 0 \\
        \end{array}
      \right).
\end{eqnarray}
It is straightforward to check that
\begin{eqnarray}
\langle\xi_{+}|\Delta \tilde{H}_{0}|\xi_{+}\rangle&=&k_{y}, \,\langle\xi_{+}|\Delta \tilde{H}_{1}|\xi_{+}\rangle=k_{y}-i\lambda,\,\nn\\
\langle\xi_{+}|\Delta \tilde{H}_{1}^{\dag}|\xi_{+}\rangle&=&k_{y}+i\lambda,
\end{eqnarray}
which lead to
\begin{eqnarray}
\left(
  \begin{array}{cc}
    \langle\tilde{\Psi}_{1}^{\rm o}|\Delta H |\tilde{\Psi}_{1}^{\rm o}\rangle & \langle\tilde{\Psi}_{1}^{\rm o}|\Delta H |\tilde{\Psi}_{2}^{\rm o}\rangle \\
    \langle\tilde{\Psi}_{2}^{\rm o}|\Delta H |\tilde{\Psi}_{1}^{\rm o}\rangle & \langle\tilde{\Psi}_{2}^{\rm o}|\Delta H |\tilde{\Psi}_{2}^{\rm o}\rangle \\
  \end{array}
\right)=\frac{3}{4}\left(
  \begin{array}{cc}
    0 & k_{x}-i\lambda  \\
    k_{x}+i\lambda   & 0 \\
  \end{array}
\right).
\end{eqnarray}
Therefore, the low-energy effective Hamiltonian for the edge state
takes the form of
\begin{eqnarray}
H_{\rm eff}=\frac{3}{4}(k_{y}\sigma_{x}+\lambda \sigma_{y}).
\end{eqnarray}

\subsection{Sample occupying $x<0$}

Finally, we study the sample occupying the $x<0$ region (Fig.\ref{boundary}d). We need to solve the eigenvalue equation:
\begin{eqnarray}
\left(
  \begin{array}{ccccccc}
    \tilde{H}_{0} & \tilde{H}_{1}^{\dag} & \tilde{H}_{2}^{\dag} & 0& 0 & 0 & \cdots\\
    \tilde{H}_{1} & \tilde{H}_{0} & \tilde{H}_{1}^{\dag} & \tilde{H}_{2}^{\dag} & 0 & 0 & \cdots \\
    \tilde{H}_{2} & \tilde{H}_{1} & \tilde{H}_{0} & \tilde{H}_{1}^{\dag} & \tilde{H}_{2}^{\dag} & 0 & \cdots \\
    0 & \tilde{H}_{2} & \tilde{H}_{1} & \tilde{H}_{0} & \tilde{H}_{1} & \tilde{H}_{2}^{\dag} & \cdots \\
    \vdots & \vdots & \vdots & \vdots & \vdots & \vdots & \ddots \\
  \end{array}
\right)\left(
         \begin{array}{c}
           \tilde{\psi}_{-1} \\
           \tilde{\psi}_{-2} \\
           \tilde{\psi}_{-3} \\
           \tilde{\psi}_{-4} \\
           \vdots \\
         \end{array}
       \right)=E\left(
         \begin{array}{c}
           \tilde{\psi}_{-1} \\
           \tilde{\psi}_{-2} \\
           \tilde{\psi}_{-3} \\
           \tilde{\psi}_{-4} \\
           \vdots \\
         \end{array}
       \right).
\end{eqnarray}
Following the same steps as in previous sections, we find
two $E=0$ modes at $k_{y}=0$ and $\lambda=0$, one of which is
$|\tilde{\Psi}_{1}'\rangle=\sum_{j}\alpha_{j}|j\rangle\otimes|\xi_{-}\rangle$,
and the other is $|\tilde{\Psi}_{2}'\rangle=\sum_{j}\beta_{j}|j\rangle\otimes|\xi_{-}\rangle$, where we continue to use $|\xi_{\pm}\rangle=(1,\pm 1)^{T}/\sqrt{2}$ to denote the two eigenvectors of $\tau_{x}$.
After orthogonalzation, the wave functions take the form of
\begin{eqnarray}
|\tilde{\Psi}_{1}^{'\rm o}\rangle&=&|\tilde{\Psi}_{1}'\rangle,\nonumber\\
|\tilde{\Psi}_{2}^{'\rm o}\rangle&=&\mathcal{N}(|\tilde{\Psi}_{2}'\rangle- |\tilde{\Psi}'_{1}\rangle\langle\tilde{\Psi}'_{1} |\tilde{\Psi}_{2}'\rangle).
\end{eqnarray}
In the neighborhood of $k_y=\lambda=0$, we expand the Hamiltonian to the first order of $k_y$ and $\lambda$: $\tilde{H}_{i}(k_{y},\lambda)=\tilde{H}_{i}(0,0)+\Delta \tilde{H}_{i}(k_{y},\lambda)$, which satisfy
\begin{eqnarray}
\langle\xi_{-}|\Delta \tilde{H}_{0}|\xi_{-}\rangle&=&-k_{y}, \,\langle\xi_{-}|\Delta \tilde{H}_{1}|\xi_{-}\rangle=-(k_{y}-i\lambda),\,\nn\\
\langle\xi_{-}|\Delta \tilde{H}_{1}^{\dag}|\xi_{-}\rangle&=&-(k_{y}+i\lambda),
\end{eqnarray}
therefore, we have
\begin{eqnarray}
\left(
  \begin{array}{cc}
    \langle\tilde{\Psi}_{1}^{'\rm o}|\Delta H |\tilde{\Psi}_{1}^{'\rm o}\rangle & \langle\tilde{\Psi}_{1}^{'\rm o}|\Delta H |\tilde{\Psi}_{2}^{'\rm o}\rangle \\
    \langle\tilde{\Psi}_{2}^{'\rm o}|\Delta H |\tilde{\Psi}_{1}^{'\rm o}\rangle & \langle\tilde{\Psi}_{2}^{'\rm o}|\Delta H |\tilde{\Psi}_{2}^{'\rm o}\rangle \\
  \end{array}
\right)=-\frac{3}{4}\left(
  \begin{array}{cc}
    0 & k_{y}+i\lambda  \\
    k_{y}-i\lambda  & 0 \\
  \end{array}
\right),
\end{eqnarray}
thus the low-energy effective Hamiltonian for the edge state
is given by
\begin{eqnarray}
H_{\rm eff}=\frac{3}{4}(-k_{y}\sigma_{x}+\lambda \sigma_{y}).
\end{eqnarray}

\end{document}